\documentclass[usenatbib]{mnras}

\usepackage{newtxtext,newtxmath}
\usepackage[T1]{fontenc}
\usepackage{ae,aecompl}

\usepackage{graphicx}	% Including figure files
\usepackage{amsmath}	% Advanced maths commands
\usepackage{amssymb}	% Extra maths symbols
\usepackage{gensymb}
\usepackage{times}
\title[Atomic cloud lifetimes in the LMC]{Towards a multi-tracer timeline of star formation in the LMC -- I.\ Deriving the lifetimes of H\,{\sc i} clouds}
\author[J. L. Ward et al.]{
Jacob L. Ward,$^{1}$\thanks{E-mail: jakelward1@gmail.com (JLW)}
M\'{e}lanie Chevance,$^{1}$ 
J. M. Diederik Kruijssen,$^{1}$ \newauthor
Alexander P. S. Hygate,$^{2,1}$  
Andreas Schruba,$^{3}$ 
and Steven N. Longmore$^{4}$ 
\\
$^{1}$Astronomisches Rechen-Institut, Zentrum f\"{u}r Astronomie der Universit\"{a}t Heidelberg, M\"{o}nchhofstra{\ss}e 12-14, 69120 Heidelberg, Germany\\
$^{2}$Max-Planck Institut f\"{u}r Astronomie, K\"{o}nigstuhl 17, 69117 Heidelberg, Germany\\
$^{3}$Max-Planck Institut f\"{u}r Extraterrestrische Physik, Giessenbachstra{\ss}e 1, 85748 Garching, Germany\\
$^{4}$Astrophysics Research Institute, Liverpool John Moores University, IC2, Liverpool Science Park, 146 Brownlow Hill, Liverpool L3 5RF,\\ United Kingdom
}

\date{Accepted XXX. Received YYY; in original form ZZZ}

\pubyear{2020}

\begin{document}
\label{firstpage}
\pagerange{\pageref{firstpage}--\pageref{lastpage}}
\maketitle

% Abstract of the paper
\begin{abstract}
	The time-scales associated with the various stages of the star formation process remain poorly constrained. This includes the earliest phases of star formation, during which molecular clouds condense out of the atomic interstellar medium. We present the first in a series of papers with the ultimate goal of compiling the first multi-tracer timeline of star formation, through a comprehensive set of evolutionary phases from atomic gas clouds to unembedded young stellar populations. In this paper, we present an empirical determination of the lifetime of atomic clouds using the Uncertainty Principle for Star Formation formalism, based on the de-correlation of H$\alpha$ and H\,{\sc i} emission as a function of spatial scale. We find an atomic gas cloud lifetime of 48$\substack{+13\\-8}$\,Myr. This timescale is consistent with the predicted average atomic cloud lifetime in the LMC (based on galactic dynamics) that is dominated by the gravitational collapse of the mid-plane ISM. We also determine the overlap time-scale for which both H\,{\sc i} and H$\alpha$ emission are present to be very short ($t_{\text{over}}<1.7$\,Myr), consistent with zero, indicating that there is a near-to-complete phase change of the gas to a molecular form in an intermediary stage between H\,{\sc i} clouds and H\,{\sc ii} regions. We utilise the time-scales derived in this work to place empirically determined limits on the time-scale of molecular cloud formation. By performing the same analysis with and without the 30 Doradus region included, we find that the most extreme star forming environment in the LMC has little effect on the measured average atomic gas cloud lifetime. By measuring the lifetime of the atomic gas clouds, we place strong constraints on the physics that drives the formation of molecular clouds and establish a solid foundation for the development of a multi-tracer timeline of star formation in the LMC.
\end{abstract}

% Select between one and six entries from the list of approved keywords.
% Don't make up new ones.

%%%%%%%%%%%%%%%%%%%%%%%%%%%%%%%%%%%%%%%%%%%%%%%%%%

%%%%%%%%%%%%%%%%% BODY OF PAPER %%%%%%%%%%%%%%%%%%

\section{Introduction}

As one of the key drivers of galaxy evolution, the mechanism by which gas is converted into stars is one of the
most fundamental processes in physics.
Observationally, star formation is traced through various phenomena, observed at a wide variety of wavelengths, and driven by the underlying physical and chemical evolution of star forming regions. 
 These range from atomic gas clumps traced by H\,{\sc i}, through molecular clouds traced in CO, the arrival of deeply embedded protostars signified by mid- to far-infrared emission,
and finally to the emergence of newly formed stars characterised by H$\alpha$ and UV emission.

The time-scales associated with these various phases of the star formation process remain poorly constrained. This is typified by (but is by no means limited to) estimates for molecular cloud lifetimes that vary from $\sim$1\,Myr in the Central Molecular Zone of the Milky Way \citep{Kruijssen2015,Henshaw2016,Barnes2017,Jeffreson2018b} to over 100\,Myr based on the presence of molecular gas within inter-arm regions of galaxies \citep{ScovilleSolomonSanders1979,Scoville1979,Koda2009}, while approaches using dynamical arguments and the classification of clouds based on stellar content yield lifetimes of a few tens of Myr \citep{Elmegreen2000,Engargiola2003,Kawamura2009,Meidt2015}.  Meanwhile, many phases of star formation have no empirically determined time-scales associated with them, including the lifetimes of atomic gas clouds from which molecular clouds condense. Without better constraints on the time-scales associated with these early stages of star formation, it is impossible to reliably model the star formation process on the cloud scale. Recent, major developments of observational facilities and novel analysis frameworks now allow this hurdle to be overcome.

The assembly times of molecular clouds from their atomic cloud progenitors have never been directly measured; however, it has been estimated  (based on the Galactic star formation rate and the Galactic gas mass) that in the Milky Way this time-scale is of the order of 30\,Myr with a longer ($\sim$50\,Myr) time-scale for the solar neighbourhood in particular \citep{Larson1994}. In a study of the transition from atomic hydrogen to molecular gas, \citet{Goldsmith2007} estimate a minimum time-scale for molecular cloud formation of 10$^{7}$\,yr. This is largely consistent with previous estimates of the ages of molecular clouds based on the formation rate estimates of molecular hydrogen (10$^{6}$--10$^{7}$\,yr, \citealt{Allen1976}). 

To obtain an empirical measurement of the time-scale for the formation of molecular clouds, it is necessary to constrain both the lifetime of their atomic cloud progenitors, and the time for which both atomic and molecular emission co-exist.
  While the classification of clouds based on their stellar content is widely implemented to obtain molecular cloud lifetime estimates in nearby galaxies, these estimates are heavily dependent on the ability to resolve cloud structures as well as the association of individual clusters or H\,{\sc ii} regions with specific clouds. The classifications used on a cloud-by-cloud basis are also highly subjective, as the definition of what constitutes a single cloud in a hierarchically structured interstellar medium is fundamentally problematic and even more so when departing from the most commonly used high-density gas tracers. Therefore, an alternative method for calculating the relative timescales of evolutionarily-linked astronomical structures that does not rely on arbitrary cloud classifications is required.
 
 A method for calculating the relative time-scales of two phases of an evolutionary sequence was proposed by \citet{KL14} and further developed in \citet{Kruijssen2018}. This \textquoteleft uncertainty principle for star formation\textquoteright method utilises the bias introduced in the measured gas-to-stellar flux ratio when focusing on gas peaks and stellar peaks to measure the relative time-scales associated with the gas and stellar phases, as well as the duration for which they coexist, and the separation length between independent regions of star formation. This formalism improves on previous methods that rely on cloud classification, because it is a priori agnostic about the definitions of clouds or H~{\sc ii} regions. Instead, it measures the lifecycle of the independent `cycling' units that together constitute a galaxy. Comparison of the properties of these units to more subjectively defined structures in molecular gas maps show that they correspond to molecular clouds in these maps \citep{Kruijssen2019,Chevance2019inprep}.
 
 The above measurement is made using the de-correlation between both types of regions as a function of spatial scale to obtain a characteristic length scale by which `independent regions' are separated. These regions reside on an evolutionary timeline at a point that is independent of their neighbours. As a result, the results obtained with this method are only weakly sensitive to the details of the identification of emission peaks. In addition, it requires only the separation length to be resolved rather than needing to resolve individual regions, as is the case for classification-based methods. This enables a broader, systematic application of this formalism to the nearby galaxy population. We refer to \citet{KL14} and \citet{Kruijssen2018} for further details.
 
 This new technique has recently been applied to determine the lifetimes of molecular clouds in NGC300 \citep{Kruijssen2019} and M33 \citep{Hygate2019inprep}. An application of the same method to characterise the lifetimes of molecular clouds as a function of environment across a sample of nine nearby, star-forming disc galaxies is presented by \citet{Chevance2019}. Because the \textquoteleft uncertainty principle\textquoteright methodology is purely statistical in nature, its applications are not limited to determining molecular cloud lifetimes. It is possible to determine the ratio of lifetimes for any pair of tracers, as long as they are evolutionarily-linked. Here, we utilise the flexible nature of this method to constrain the atomic gas cloud lifetime.

The Large Magellanic Cloud (LMC) is the largest and most massive of the Milky Way's satellites. It is a gas-rich, star-forming, dwarf-irregular galaxy with a close-to-face-on inclination (25.7$\degr$, \citealt{Subramanian2013}), a modest distance ($\sim$50\,kpc,\,\citealt{Laney2012}), and hosts the well known starbursting region 30 Doradus. Therefore, the LMC presents a unique opportunity for multi-scale star formation studies, allowing observations from the scale of an entire galaxy, down to the sub-pc circumstellar environments of individual stars and protostars.  For the current project, our key interest in the LMC over other nearby galaxies is that it is fully sampled at a wider range of wavelengths than any other galaxy in the Universe. This unique feature of the LMC allows us to utilise the wide range of survey data available and develop the first multi-tracer timeline of star formation. In doing so, we will determine several of the key remaining unknowns of star formation models, including cloud assembly times, cloud lifetimes, and the \textquoteleft overlap\textquoteright\,time-scales during which different star formation tracers co-exist. We also aim to derive additional properties such as the star formation efficiency (SFE), the typical stellar feedback velocity and the fraction of energy deposited by feedback that acts on the ISM.

This paper is the first in a series with the aim of realising this attempt to develop a multi-tracer timeline of star formation. Here we present a measurement of the lifetimes of the earliest structures in the star formation process: clouds of atomic gas traced by neutral hydrogen.
 In Section \ref{obsdat}, we outline the data used to achieve this goal. In Section \ref{UPsec}, we briefly introduce the \textquoteleft Uncertainty Principle\textquoteright\, methodology that we apply in this work.
 The derived time-scales are presented in Section \ref{results} as well as the relevant byproducts and subsequently derived parameters. We discuss the robustness of these results and their implications within the broader context
of star formation in the LMC in Section \ref{discussion}. In Section \ref{conclusions}, we summarise the key findings of this paper.

\section{Observations}

\label{obsdat}
Our methodology is used to determine the {\it relative} duration of two phases of star formation.
Therefore, in order to determine the {\it absolute} characteristic time-scale of H\,{\sc i} overdensities, H$\alpha$ emission is used to provide a \textquoteleft reference\textquoteright\, time-scale.
The H$\alpha$ emission line map used in this work is the mosaic produced by the Magellanic Clouds Emission Line Survey (MCELS, \citealt{Smith1998,Smith2005}). This map includes both H$\alpha$ emission and continuum emission. The MCELS H$\alpha$ filter is centred at 6563\,\AA{}, with a full-width at half-maximum of 30\,\AA{}. In Appendix \ref{Appendix_A}, we make use of the continuum-subtracted H$\alpha$ image of the Southern H-Alpha Sky Survey Atlas (SHASSA; \citealt{Gaustad2001}) to test that consistent results are obtained using both images as reference maps. The large ratio between H$\alpha$ emission and H\,{\sc i} emission time-scales makes the H$\alpha$ plus continuum emission map from MCELS a better choice than the continuum subtracted map for the present study. This is discussed in Section \ref{robostness}.

One of the early steps in the process of applying the \textquoteleft Uncertainty Principle\textquoteright\, framework is the identification of emission peaks (see Section \ref{UPsec}).
The presence of an existing catalogue of H\,{\sc ii} regions in the LMC based on the MCELS H$\alpha$ map \citep{Pellegrini2012} provides a pre-defined upper limit ($n_{\text{H}\alpha} < $\, 401) on the number of independent recent massive star formation events in the LMC, which can be referred to in order to check the validity of the detected emission peaks.  The \citet{Kim2003} H\,{\sc i} map of the LMC obtained with Parkes and ATCA is used as a tracer of neutral atomic gas in the LMC.  We can expect on the order of 1000 clouds in the H\,{\sc i} emission map \citep{Kim2007}, as we make use of the same data. These pre-defined upper limits on the number of independent emission peaks in both tracers act as benchmarks for the peak-selection process.

As well as providing benchmarks for peak selection, quantities drawn from previous studies of the LMC can be used to calculate composite parameters such as the star formation efficiency (see Section \ref{compos_Sct}).
The total H\,{\sc i} mass of the LMC derived by \citet{Bruns2005} and \citet{Staveley-Smith2003} are M$_{\text{H\,{\sc i}}} = (4.41 \pm 0.09) \times 10^{8}$\,M$_{\sun}$ and M$_{\text{H\,{\sc i}}} = (4.8 \pm 0.2) \times 10^{8}$\,M$_{\sun}$, respectively.
Current estimates of the star formation rate (SFR) in the LMC fall in the range
0.05--0.25\,M$_{\sun}$yr$^{-1}$ \citep{Whitney2008}.

\section{The uncertainty principle for star formation method}

\label{UPsec}

The \textquoteleft uncertainty principle for star formation\textquoteright, introduced by \citet{KL14}, describes the breakdown of star formation relations at small spatial scales \citep{Onodera2010,Schruba2010,Liu2011} as a result of incomplete sampling of star forming regions. Assuming a timeline consisting of a gas phase, followed by a young stellar phase, with a time-scale over which both phases co-exist, the measured gas depletion time within an aperture focused on a gas peak or a stellar peak is biased towards larger or smaller values, respectively. When successively smaller apertures are focused on gas peaks and stellar peaks, the bias in the measured gas-to-stellar flux increases with respect to the galactic average. This behaviour was first noted by \citet{Schruba2010} and can be reproduced by a characteristic tuning-fork-shaped, non-degenerate model with three free parameters: the relative time-scale between the stellar and gas phase, the overlap time-scale during which both phases co-exist, and the characteristic separation length between independent star forming regions. One key advantage of this technique, therefore, is that it does not require star forming regions to be resolved, only the separations between star forming regions. Details on how the most simple incarnation of this model is derived can be found in \citet{KL14}.

 While this framework is usually discussed in the context of a gas tracer and a star formation tracer as above, the purely statistical nature of the Uncertainty Principle means that it can be applied to any two tracers which are evolutionarily-linked such that one tracer will be followed by the other tracer in each independent region in the future. This opens up the possibility of deriving a multi-tracer time-scale of star formation. In this paper we derive the time-scale over which atomic gas clouds traced by H\,{\sc i} evolve into young massive star forming regions probed by H$\alpha$. We use the known duration of the H$\alpha$ emsision as a reference to calibrate the relative timeline derived from our model.

\subsection{The H{\sc eisenberg} code}

The Uncertainty Principle has been implemented in this work using the H{\sc eisenberg} code, using two input images of the same galaxy in tracers that are evolutionarily-linked (in this case H\,{\sc i} emission and H$\alpha$ emission) in order to calculate the relative time-scales associated with the two maps. Using a known reference time-scale for one of the two images allows the absolute time-scale associated with the second image to be derived. We refer to the image for which we determine a time-scale as the {\it target map} and the image associated with the reference time-scale as the {\it reference map}. Here we present a brief, qualitative description of the key functions of this code, as described in detail in \citet{Kruijssen2018}.

First, the input images for each tracer are regridded by convolving them to the best common spatial resolution. At this stage, the inclination of the galaxy is also taken into account in the regridding process. We have adopted values for the inclination and line-of-nodes position angle for the LMC of 25.7\textdegree\, and 141.5\textdegree, respectively, from \citet{Subramanian2013}. Then, if necessary, masks are read in the form of {\sc DS9}\footnote{\url{http://ds9.si.edu.}} region files and applied to the images. These masks are used to restrict the analysis to specific regions within the galaxy. In this work we mask pixels beyond a radius of 3.8\textdegree\, ($\sim$3.3\,kpc) from the centre of the MCELS H$\alpha$ image. This is to ensure that the maps completely overlap in their coverage, and that we focus on the star-forming part of the LMC rather than the quiescent outskirts. Later in this work (Section \ref{30Dorsec}), we use masks to investigate the effect that the extreme star-forming region 30 Doradus has on the mean time-scales derived in this work.

The images are then convolved to a range of spatial scales, with the aim of measuring the flux within a given radius around each flux peak in the images. We consider a minimum aperture size of 25\,pc and a maximum aperture size of 3200\,pc, sampled in logarithmic steps with 12 aperture sizes in total. A tophat kernel is used to convolve the images to each desired spatial scale. 
Emission peaks are then identified in both maps using {\sc clumpfind}\footnote{\url{http://www.ifa.hawaii.edu/users/jpw/clumpfind.shtml}}\,\citep{Williams1994}. The fluxes within apertures of various sizes as specified in the input parameter file (the same as the spatial scales that the images are previously convolved to) are then measured. 

The Uncertainty Principle formalism assumes that the identified peaks reasonably sample the underlying timeline \citep{Kruijssen2018}. This is a much weaker assumption than requiring the identified peaks to homogeneously sample the complete evolutionary timeline, which would imply that relative peak counts could be used as a direct proxy for the relative time-scales, but additionally would make the method strongly dependent on the (subjective) peak identification process. By contrast, the formalism adopted here avoids this problem by using the identified peaks only to centre the apertures within which the gas-to-stellar flux ratio is measured on gas or stellar flux overdensities. As a result, the details of the peak identification do not affect strongly the derived quantities, as long as an \textquoteleft obvious\textquoteright set of peaks is identified \citep{Kruijssen2018}. In addition, we note that these requirements do not imply that the emission peaks detected using {\sc clumpfind} correspond to independent star-forming regions (and the method does not assume so).

For each aperture size, the gas-to-stellar flux ratio when focusing on peaks of each tracer is calculated and is normalised with respect to the galactic average flux ratio between the two tracers. Independence-weighted uncertainties are then calculated for all of these values. Then, the characteristic two-branch tuning-fork shaped model is fitted to the observed bias values to constrain the evolutionary timeline.  A reduced-$\chi^{2}$ fit of the model to the data points (see Section~\ref{results}) and their uncertainties is carried out using three free parameters: the relative time-scales of the two tracers, the overlap time for which the two tracers co-exist, and the characteristic separation length. This returns not only the best-fit values but also the three-dimensional probability distribution function (PDF) of the free parameters. The marginalised one-dimensional PDFs for each of the free parameters are obtained (see Section~\ref{results}) as well as the two-dimensional PDF for each parameter pair. These three empirically determined parameters are then used to calculate a range of derived quantities. See table 4 of \citet{Kruijssen2018} for a full list of derived quantities.

The robustness of the results obtained with H{\sc eisenberg} can be assessed using a number of criteria outlined in \cite{Kruijssen2018}. The criteria relevant to the present study are discussed in detail in Section \ref{robostness}, and are summarised here:
\begin{enumerate}    
\item The durations of the reference and target phases in question must not differ by more than an order of magnitude.
\item The characteristic length scale by which `independent regions' are separated must be larger than the effective resolution of the smallest aperture size.
\item The potential precision of the derived time-scales is dependent on the number of peaks identified.
\item The SFR should not vary by more than 0.2\,dex over the reference or derived time-scales.
\item Each independent region should be visible in both tracers at some point in its lifecycle.
\item In order to obtain a measurement of the time-scale for which the two tracers co-exist, $t_{\text{over}}$, rather than placing a limit on $t_{\text{over}}$, we require that $0.05 < t_{\text{over}}/\tau < 0.95$, where $\tau$ is the total time-scale of the evolutionary process in question, i.e.\ the integrated time-scale over which either tracer is visible (or both of them are).
\end{enumerate}

\subsection{Iterative Fourier filtering}

\label{fourier}

In addition to emission originating from the star-forming (and future star-forming) regions themselves, both H$\alpha$ and H\,{\sc i} emission exhibit significant diffuse components. As these diffuse components are not directly associated with any one star forming region, they do not directly participate in the cycle of star formation between H\,{\sc i} overdensities and young star forming regions. Therefore they are considered to be contaminants in the context of this study. The input images are filtered using a Gaussian-shaped filter in Fourier space to remove the low-spatial-frequency (large-scale), diffuse emission component from the images. This Fourier filtering is applied in an iterative fashion: the uncertainty principle method is applied to the data, the value of the characteristic distance between independent regions ($\lambda$) is then multiplied by a conversion factor to determine the scale on which emission is cut in Fourier space. The Heisenberg code is then applied to the resultant filtered images and a new value of lambda is measured, which determines a new filter in Fourier space, which is then applied to the unfiltered images. This process is repeated until the derived parameters are consistent with those of the three previous iterations to within 5 per cent.
The application of these filters in Fourier space is described in detail in \citet{Hygate2018}.

\begin{figure*}
	    \includegraphics[width=0.47\linewidth]{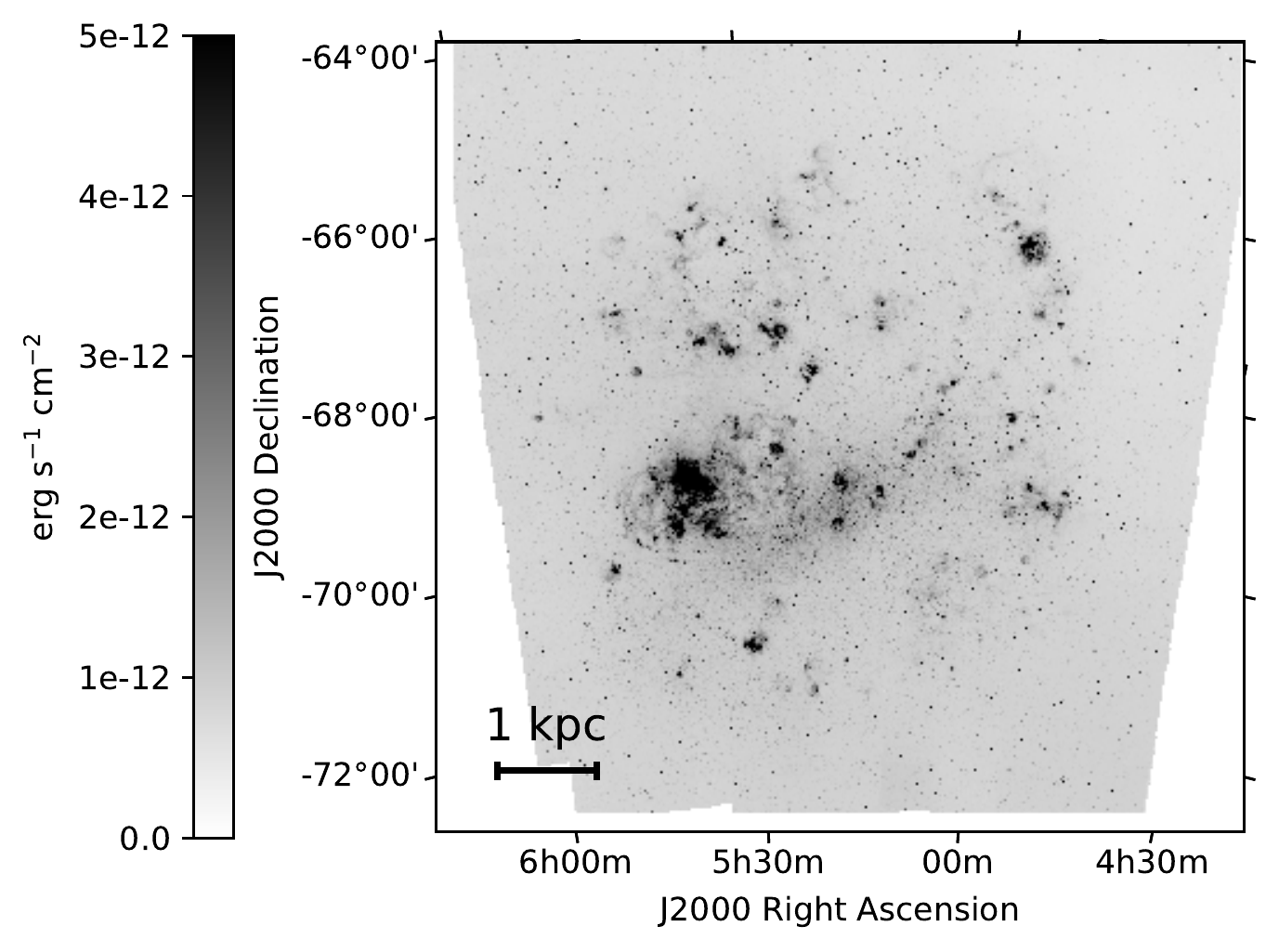}
	     \includegraphics[width=0.47\linewidth]{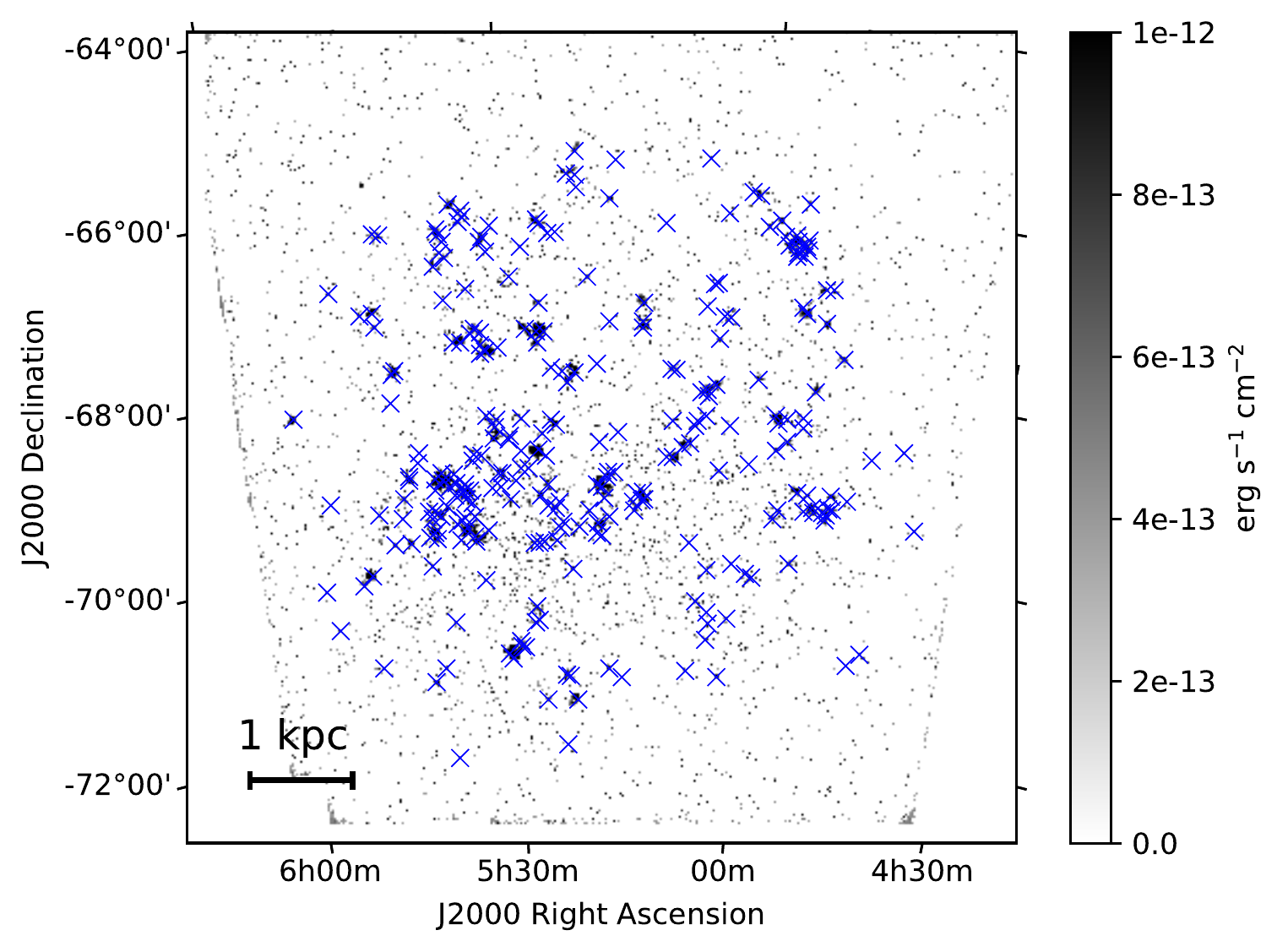}
	      \includegraphics[width=0.49\linewidth]{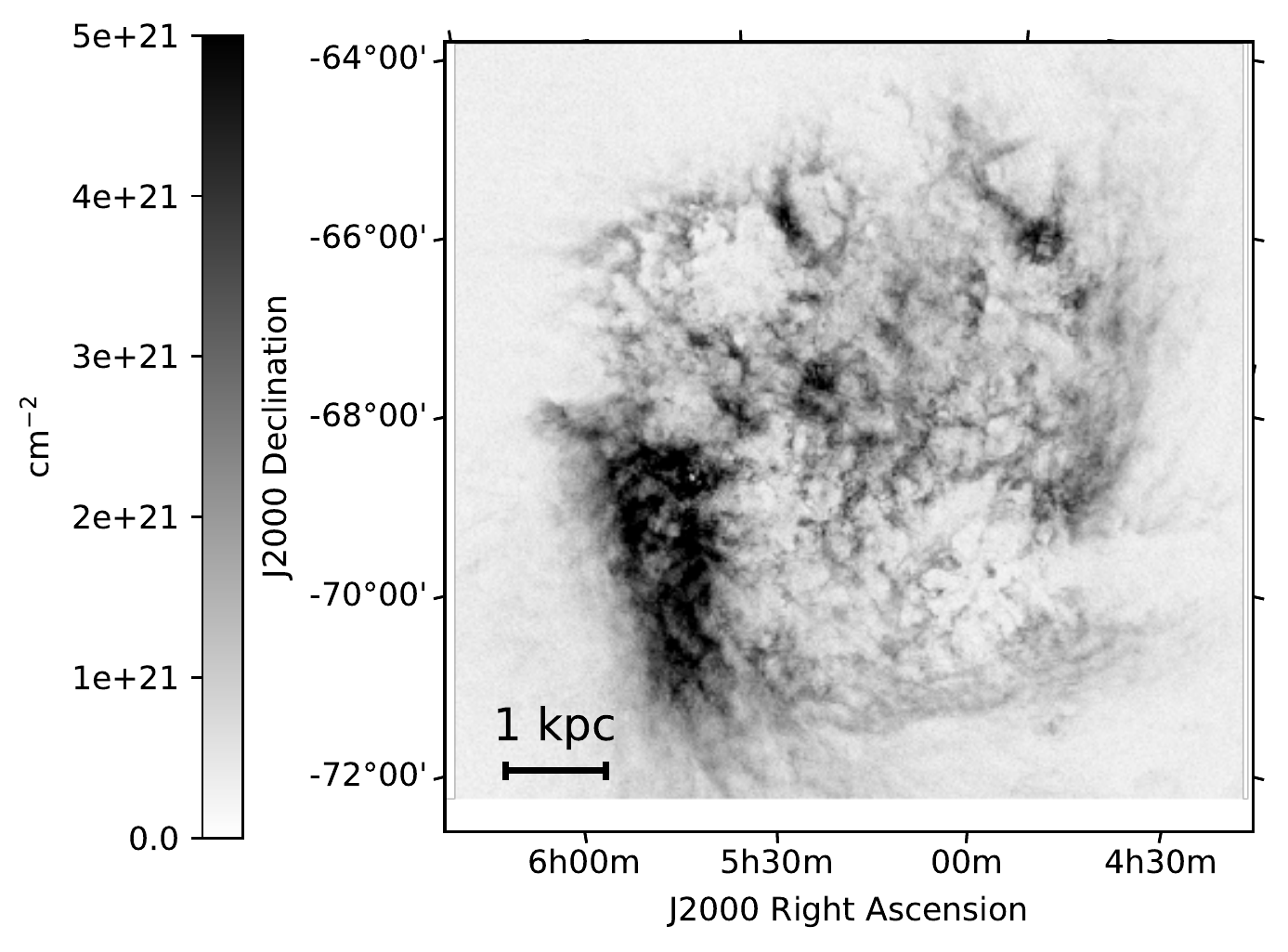}
	       \includegraphics[width=0.50\linewidth]{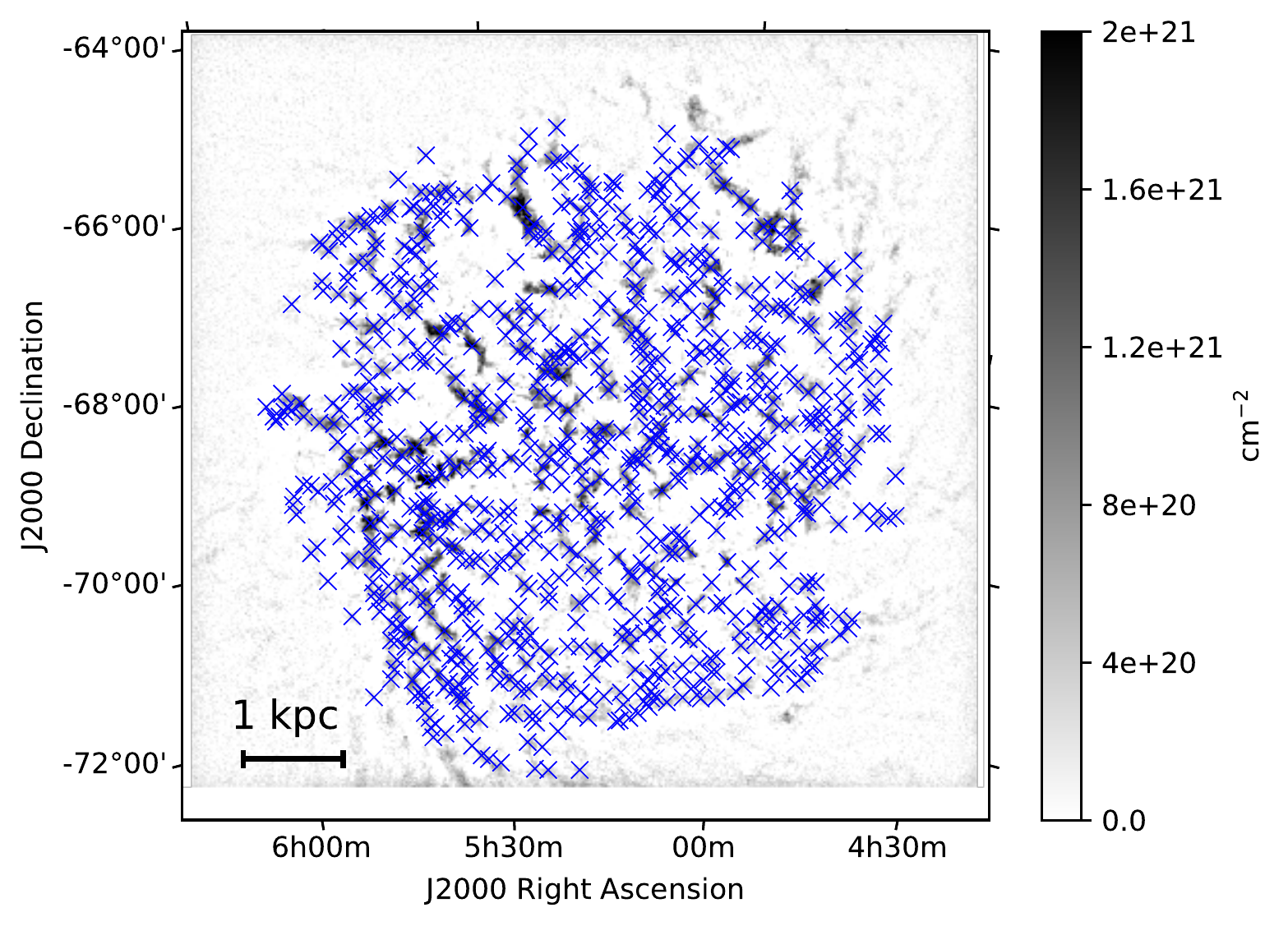}
		\caption{\label{example_filtering} Images of the LMC before (left) and after (right) the iterative Fourier filtering process in  H$\alpha$ {plus continuum emission} (top) and H\,{\sc i} column density (bottom). In the filtered images (right-hand panels), the emission peaks selected by the {\sc Heisenberg} code are marked by blue crosses. }
\end{figure*}

In Fig. \ref{example_filtering} we show the original emission maps and those resulting from the iterative Fourier filtering process. It is clear that the vast majority of the diffuse emission in both the MCELS  H$\alpha$ mosaic and the H\,{\sc i} emission map has been successfully removed. 89\% of the total H\,{\sc i} flux is considered to be diffuse and 87 per cent of the H$\alpha$ plus continuum emission is considered to be diffuse. \citet{Hygate2018} and \citet{Hygate2019inprep} presented two correction factors that can be calculated in order to take into account the excess flux removed from peaks as a result of the use of a Gaussian filter ($q_{\text{con}}$) and the result of overlapping regions ($q_{\text{overlap}}$). We calculate correction factors of $q_{\text{con,H}\alpha} = 0.98$ and $q_{\text{con,H{\sc i}}} = 0.96$, for the H$\alpha$ and the H\,{\sc i} images, respectively. The corresponding correction factors due to overlapping regions for the H$\alpha$ and H\,{\sc i} emission maps are $q_{\text{overlap,H}\alpha} = 0.94$ and $q_{\text{overlap,H}\alpha} = 0.87$, respectively. When these correction factors are applied, we establish diffuse fractions of $\sim$86\, per cent and $\sim$87\, per cent for the H$\alpha$ and H\,{\sc i} maps, respectively, following \citet{Hygate2018,Hygate2019inprep}.
	
It should be noted that much of the diffuse emission in the MCELS H$\alpha$ emission is made up of continuum source structures such as the bar\footnote{When using an H$\alpha$ map without continuum emission (see Appendix \ref{Appendix_A}), we obtain a diffuse emission fraction of 65 per cent.}. In this work, we have consistently used a conversion factor between the measured values of $\lambda$ and the Fourier filter cutting length of $n_{\lambda}=$10, i.e. the maps are filtered at a scale corresponding to 10 times the characteristic separation length $\lambda$. This value is chosen in order to satisfy the recommended criterion of $q_{\text{con}} > 0.9$ of \citet{Hygate2018}. Six iterations of H{\sc eisenberg} with filtering were required in order for the results to converge within a 5\% tolerance with the two previous measurements.

\subsection{Characteristic H$\alpha$ time-scales}

\label{DanTimes}

The Uncertainty Principle derives time-scales relative to the characteristic time-scale of the reference map that is used. In order to determine the absolute value of a time-scale, the average time-scale associated with the reference map must be known. The H$\alpha$ emission of a star-forming galaxy is dominated by the ionisation of gas by young stellar systems. The ionising photons primarily originate from the most massive stars and the H$\alpha$ emission is most concentrated in H\,{\sc ii} regions. \citet{Haydon2018} provide expressions derived from stellar population synthesis models to determine the characteristic time-scales associated with continuum-subtracted H$\alpha$ emission with and without continuum emission for varying metallicities and filter widths.
The characteristic time-scale associated with H$\alpha$ emission with a well-sampled initial mass function can be estimated using
\begin{equation}
    \label{Haminus_t}
    t_{\text{H}\alpha -} [\text{Myr}] = \left(4.32\substack{+0.09\\-0.23}\right) \left(\frac{Z}{Z_{\sun}}\right)^{\left(-0.086\substack{+0.010\\-0.023}\right)} \text{,}
\end{equation}
 and for an image containing both H$\alpha$ and continuum emission using
\begin{equation}
\label{haplus_eqn}
\begin{aligned}
t_{\text{H}\alpha +} [\text{Myr}] = \left(8.98 \substack{+0.40 \\ -0.50}\right) W_{0}^{\left(0.265\substack{+0.028 \\ -0.051}\right)} + \left(0.23\substack{+0.15 \\ -0.11}\right)Z_{0}W_{0} \\
-\left(0.66\substack{+0.12 \\ -0.19}\right)Z_{0} + \left(0.55\substack{+0.46 \\ -0.29}\right)W_{0}
\end{aligned}
\end{equation}
 where $W_{0}$ is the filter width in units of 40\AA{}, $W_{0} = W / 40$\AA{}, and $Z_{0}$ is the metallicity in units of solar metalicity.
 Assuming a characteristic metallicity of 0.4$Z_{\sun}$, consistent with that of the LMC \citep{Dufour1982,Bernard2008}, equation \ref{Haminus_t} yields a time-scale of $t_{\text{H}\alpha-}=4.67\substack{+0.15\\-0.34}$\,Myr for continuum-subtracted H$\alpha$ emission in the LMC. However, while continuum-subtracted H$\alpha$ emission images of the LMC are available, the low time-scale associated with these maps renders them unsuitable for deriving the much longer time-scale associated with H\,{\sc i} emission (see Section \ref{robostness}), because the applied methodology performs better when the pair of tracers used have more similar lifetimes \citep{Kruijssen2018}. For H$\alpha$ emission with a contribution from continuum emission, we derive a time-scale of 8.54$\substack{+0.97\\-0.82}$\,Myr using the same metallicity, and a filter width of 30\AA{}, consistent with that of the MCELS H$\alpha$ image. In Appendix \ref{Appendix_A}, we discuss the robustness of the time-scales derived using the results of \citet{Haydon2018}, using the SHASSA narrow band image and the MCELS H$\alpha$  mosaic of the LMC. We confirm the consistency between the continuum subtracted and non-continuum-subtracted H$\alpha$ images, thereby observationally validating the reference timescales obtained by \citet{Haydon2018}, as well as showing that the process of applying the Uncertainty Principle to observational data is both commutative and transitive, such that for a set of evolutionarily-linked time-scales $t_0$, $t_1$, and $t_2$, $t_2/t_0 = (t_1/t_0) \times (t_2/t_1)$.
 
 \section{Results}
\label{results}

Emission peaks were identified in both the H$\alpha$ and H\,{\sc i} images at a spatial scale of 25\,pc (corresponding to 103\arcsec) using logarithmically spaced sensitivity contours over a range covering 2.5\,dex (H$\alpha$) and 1.0\,dex (H\,{\sc i}) using 0.25\,dex steps. A minimum of 15 pixels is required to define an emission peak, in order to exclude point sources from the analysis.
The positions of all of the identified peaks are marked on the respective filtered images in Fig.~\ref{example_filtering}. The number of peaks selected from this process is 275 and 893 for the H$\alpha$ and H\,{\sc i} images, respectively (see also Table \ref{Fun_tab}). This is consistent with the upper limit on the number of independent H\,{\sc ii} regions set by \citet{Pellegrini2012}. The number of detected H\,{\sc i} emitting peaks is also of the order of that expected based on the analysis of \citet{Kim2003}.

\begin{figure*}
\begin{minipage}{175mm}
\includegraphics[width=0.495\linewidth]{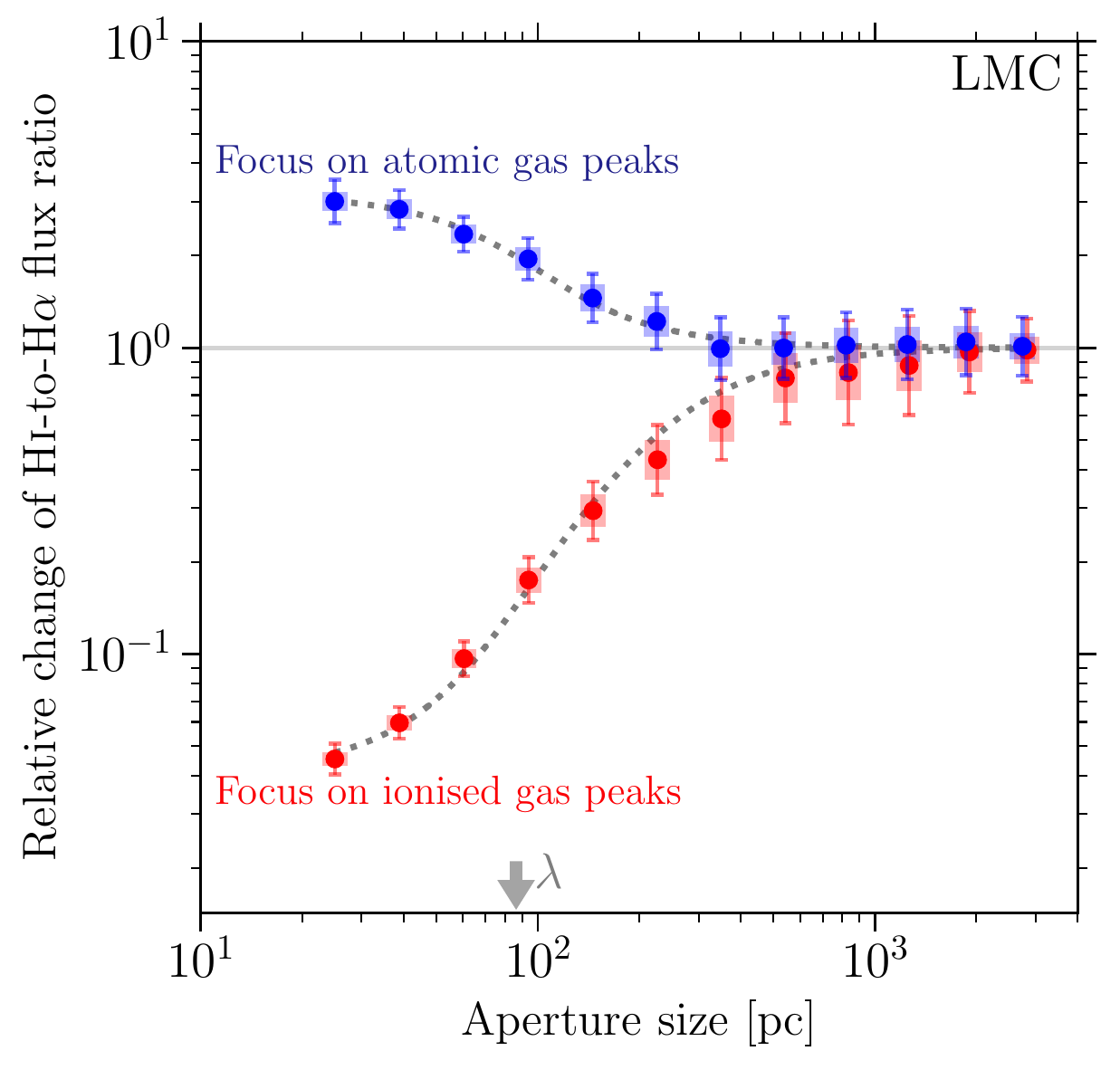}
\includegraphics[width=0.495\linewidth]{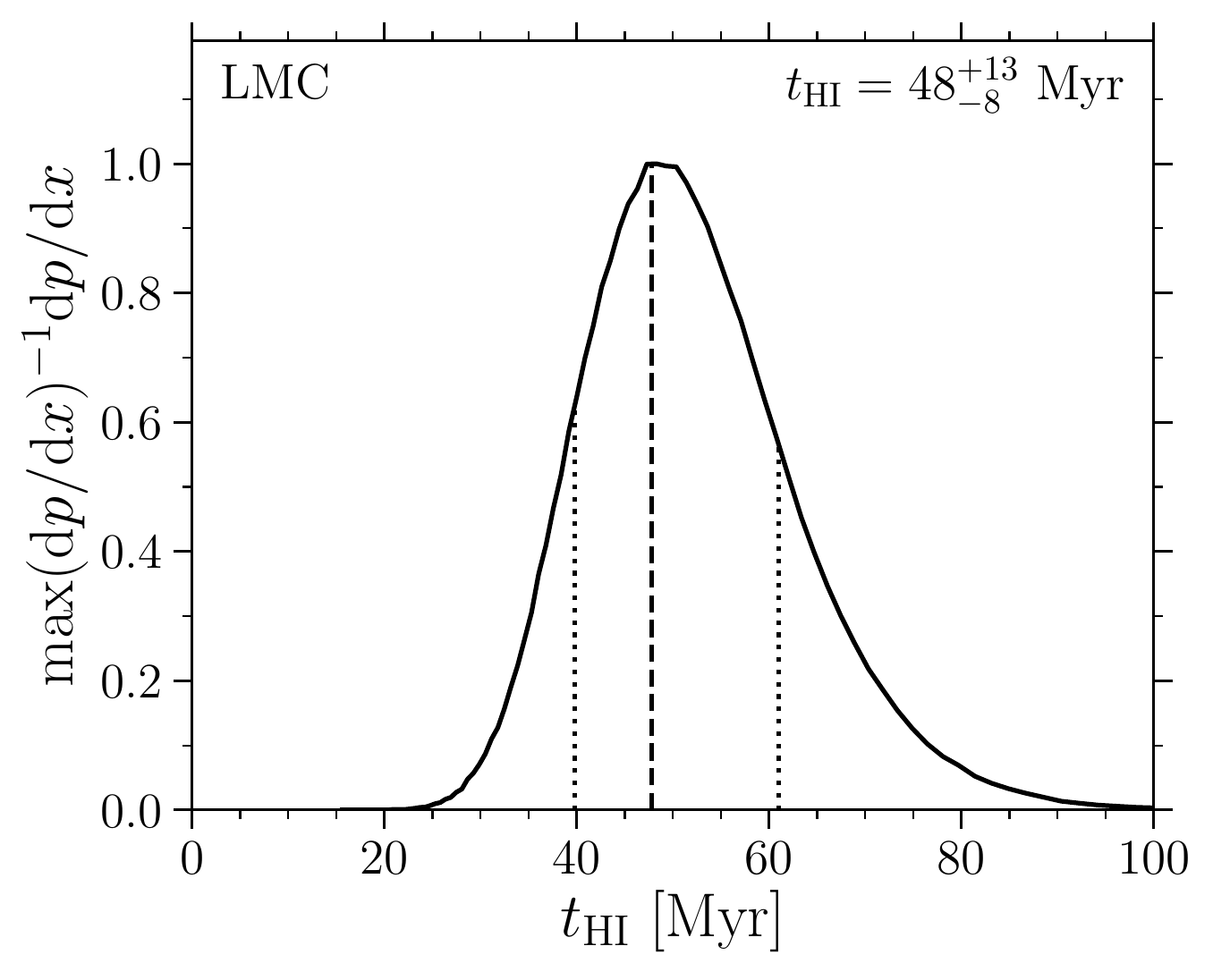}
\includegraphics[width=0.495\linewidth]{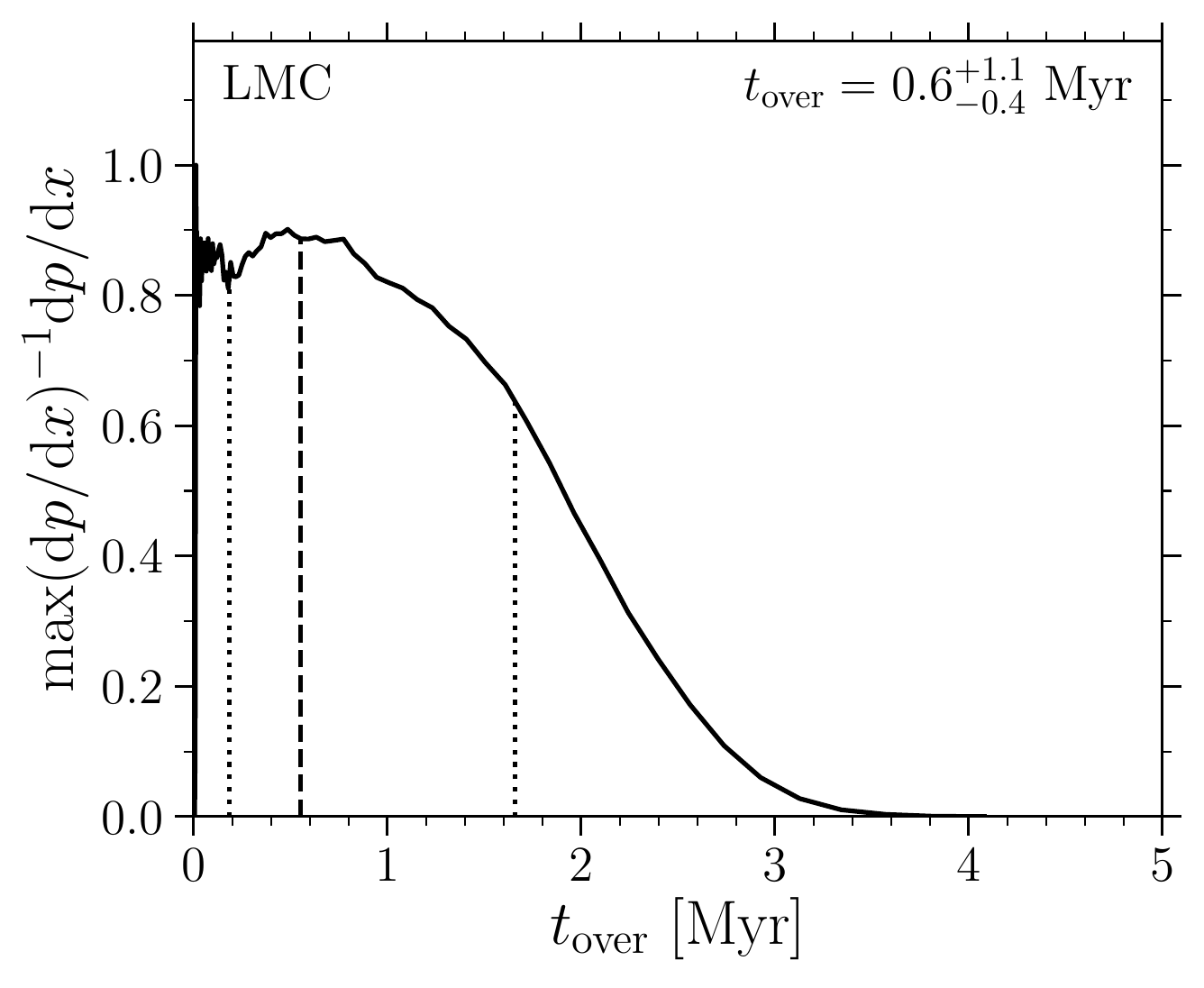}
\includegraphics[width=0.495\linewidth]{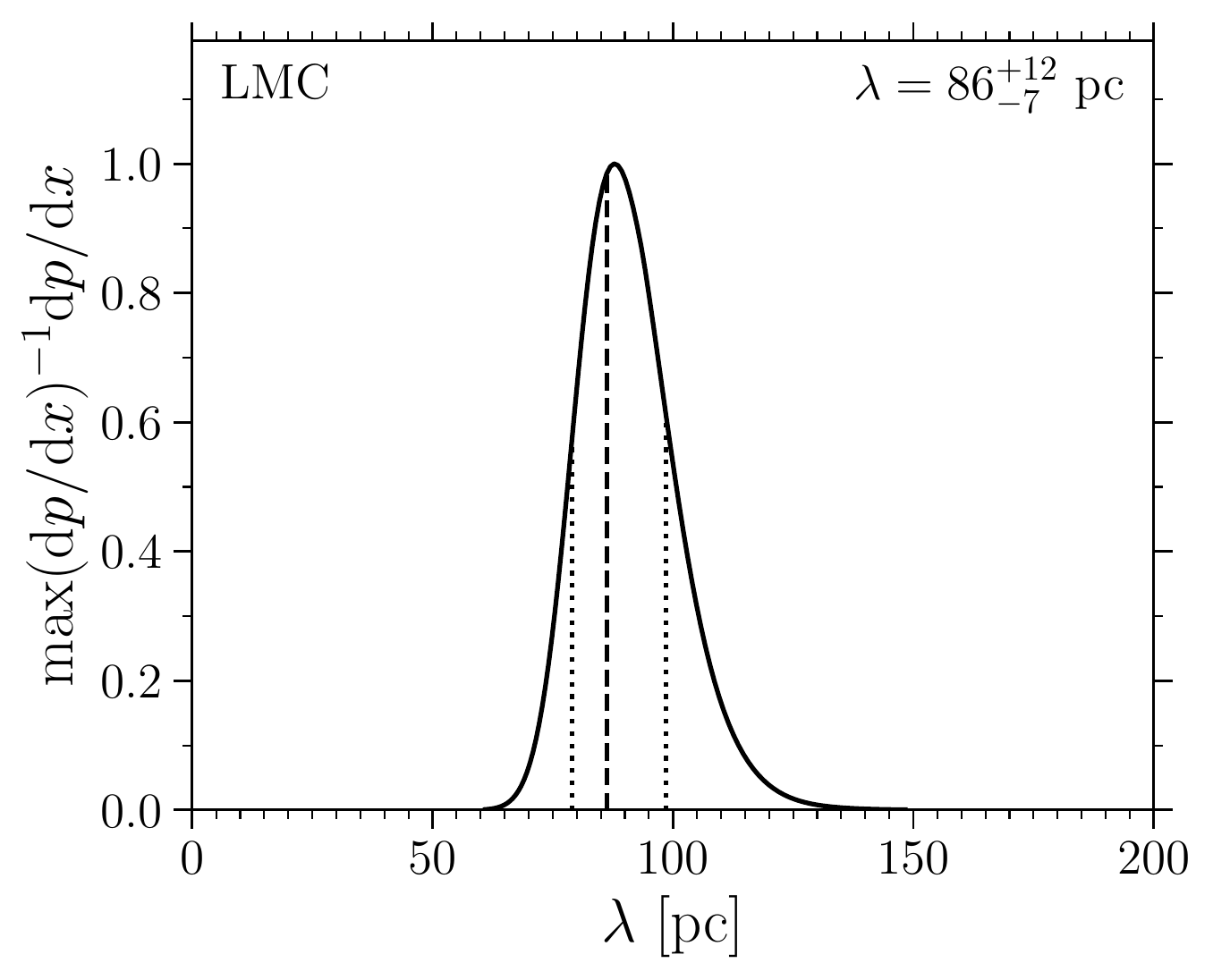}
\caption{\label{Tuningfork} Upper-left: The relative change of the gas to stellar flux ratio compared to the galactic average as a function of the aperture size when focusing on the H$\alpha$ peaks (red) and the H\,{\sc i} peaks (blue) with the fitted model shown in green. The error bars indicate the $1\sigma$ uncertainty on each individual data point, whereas the shaded areas indicate the effective $1\sigma$ uncertainty range that accounts for the covariance between the data points and should be used when visually assessing the quality of the fit. Upper-right: The PDF derived for $t_{\text{H\,{\sc i}}}$. Lower-left: The derived PDF for the overlap time-scale, $t_{\text{over}}$. Lower-right: The derived PDF for the characteristic separation length between independent regions, $\lambda$.}
 \end{minipage}
\end{figure*}

The upper-left panel of Fig. \ref{Tuningfork} shows the relative change of the gas-to-SFR tracer flux ratio compared to the galactic average versus aperture size focusing on the H$\alpha$ peaks (red) and the H\,{\sc i} peaks (blue). Shown in green is the tuning fork-shaped model fitted to the gas-to-SFR tracer flux ratios. A reduced $\chi^{2}$ value of 0.37 is obtained for the fit presented in Table \ref{Fun_tab}. 
As described in \citet{KL14} and \citet{Kruijssen2018}, the fitted model is dependent only on three fundamental parameters: the time-scale associated with the target map in units of the reference map ($t$), the relative time-scale for which both emission is present ($t_{\text{over}}$), and the characteristic distance between independent regions ($\lambda$). We now turn to a detailed description of these quantities.

\subsection{Fundamental free parameters}

In Table \ref{Fun_tab}, we present the results of fitting the Uncertainty Principle model to the filtered H$\alpha$ and H\,{\sc i} images of the LMC. In this paper, we use the MCELS H$\alpha$ plus continuum emission map of the LMC as a reference map in order to constrain the time-scale associated with H\,{\sc i} clouds.
 Figure \ref{Tuningfork} shows the PDFs for the three free parameters ($t$, $t_{\text{over}}$, and $\lambda$)\, using the MCELS H$\alpha$ plus continuum map as a reference and the H\,{\sc i} column density image as the target map. 
 In Appendix \ref{Appendix_A}, the method has been applied using continuum subtracted H$\alpha$ emission image as well as in reverse, i.e. the H\,{\sc i} image has been used as a reference map to determine the time-scale associated with the H$\alpha$ map. The use of H{\sc eisenberg} to determine absolute time-scales relies on the technique being both transitive and commutative, such that a time-scale determined with H{\sc eisenberg} can be used as reference time-scales in subsequent H{\sc eisenberg} runs. The experiments in Appendix \ref{Appendix_A} allow us to confirm that the technique is commutative and transitive and that we are therefore justified in using the characteristic time-scales introduced in Section \ref{DanTimes}.
 
\begin{table*}
\caption{\label{Fun_tab} \label{Fun_tab_abs} Fundamental parameters determined by applying H{\sc eisenberg} to the H$\alpha$ and H\,{\sc i} maps of the LMC. In columns 3 and 4, we present the number of peaks selected in each map. The three fundamental parameters fitted to the observed data: the time-scale of the target tracer ($t$), the overlapping time-scale between the two tracers ($t_{\text{over}}$), and the characteristic region separation length ($\lambda$), are presented in columns 6, 7, and 8, respectively. The reduced $\chi ^{2}$ goodness-of-fit parameter is shown in the final column. The time-scales presented in the first row are relative to the reference time-scale ($t_{\text{ref}}$) and in the second row are calibrated based on the work of \citet{Haydon2018}.}
\begin{center}
		\begin{tabular}{c c c c c c c c c}
				\hline
			reference map & target map & n$_{\text{ref}}$ & n$_{\text{target}}$ & $t_{\text{ref}}$ &  $t$ & $t_{\text{over}}$ & $\lambda$ [pc] & $\chi^{2}$ \\
			\hline
			H$\alpha$ & H\,{\sc i} & 275 & 893 & 1.0$\pm$0.0\,$t_{\text{H}\alpha}$ & 5.6$\substack{+1.4\\-0.8}$\,$t_{\text{H}\alpha}$ & $<$0.18\,$t_{\text{H}\alpha}$ & 86$\substack{+12\\-7}$ & 0.37 \\
			H$\alpha$ & H\,{\sc i} & 275 & 893 & 8.5$\substack{+1.0\\-0.8}$\,Myr & 48$\substack{+13\\-8}$\,Myr & $<$1.7\,Myr & 86$\substack{+12\\-7}$ & 0.37 \\
			\hline
		\end{tabular}
\end{center}
\end{table*}

For the LMC, we determine that the H\,{\sc i} cloud lifetime is 5.6$\substack{+1.4\\-0.8}$ times the reference time-scale associated with the MCELS H$\alpha$ image of the LMC. The derived overlap time-scale is very short, $t_{\text{over}} = 0.06\substack{+0.12\\-0.04}$\,$t_{\text{H}\alpha}$. The second row of Table \ref{Fun_tab_abs} shows the results of H{\sc eisenberg} using the reference time-scale for H$\alpha$ plus continuum emission ($8.5\substack{+1.0\\-0.8}$\,Myr) derived following \citet{Haydon2018}. When this reference time-scale is used, we determine the mean lifetime of H\,{\sc i} peaks to be 48$\substack{+13\\-8}$\,Myr. The overlap time-scale, the mean time for which both tracers are present in the same region, is determined to be 0.6$\substack{+1.1\\-0.4}$\,Myr. However, in order to obtain a value for $t_{\text{over}}$, rather than an upper limit, we require that $t_{\text{over}} > 0.05\tau$ \citep{Kruijssen2018}. As this condition is not met, we adopt an upper limit on the overlap time-scale of $t_{\text{over}} < 1.7$\,Myr.

The characteristic separation length between independent regions ($\lambda$) is derived as 86$\substack{+12\\-7}$\,pc. This characteristic region separation length coincides with the slight steepening of the power spectra at high spatial frequency observed by \citet{Elmegreen2001}. It is suggested by \citet{Elmegreen2001}, that this steepening of the power spectrum could represent a probe of the cool cloud scale-height of the disc, where the full thickness of the disc is twice this number, consistent with that derived for the LMC by \citet{Kim1999} of $\sim$180\,pc. This behaviour is strongly indicative of an ISM structure that is largely driven by feedback \citep{Kruijssen2019}, as bubbles are expected to depressurise on expansion beyond the scale height of the disc \citep{McKee1977,Hopkins2012}. This similarity between the characteristic separation length $\lambda$ and the scale height of the disc has been shown to be present in NGC 300 \citep{Kruijssen2019}.

\subsection{Byproducts and derived quantities}
\label{byproducts_etc}

In addition to the fundamental parameters derived through our methodology, there are a number of additional properties that can be derived from these. The byproducts and derived properties relevant to the current work are summarised in Table \ref{tab:byproducts}.

\begin{table}
\caption{The additional parameters for each of the three H{\sc eisenberg} runs presented in this work. The total time-scale ($\tau = t + t_{\text{ref}}$) is presented in the first row. Note that this time-scale omits any phase during which the region does not emit in either H {\sc i} or H$\alpha$ (e.g.\ a purely molecular evolutionary stage without massive star formation).
	 The average radii for H$\alpha$ emission peaks ($r_{\text{H}\alpha}$) and H\,{\sc i} peaks ($r_{\text{H\,{\sc i}}}$) are given in the second and third rows. The final two rows show the peak concentration parameters $\zeta$ (described in Section \ref{byproducts_etc}) for both H$\alpha$ and H\,{\sc i} emission. The original run is presented in the first column, and the runs in which 30 Doradus is masked (See Section \ref{30Dorsec}) are shown in the second and third columns.}
    \label{tab:byproducts}
    \begin{center}
    \begin{tabular}{l|c|c|c}
        \hline
        parameter & no mask & 200\,pc mask & 1\,kpc mask \\
        \hline
        $\tau$ [Myr] & 56$\substack{+13\\-9}$ & 58$\substack{+12\\-9}$ & 46$\substack{+8\\-7}$ \\
        $r_{\text{H}\alpha}$ [pc] & 14.9$\substack{+0.4\\-0.2}$ & 15.0$\substack{+0.4\\-0.3}$ & 15.5$\substack{+0.3\\-0.2}$\\
        $r_{\text{H\,{\sc i}}}$ [pc] & 21.5$\substack{+1.7\\-1.2}$ & 21.6$\substack{+1.3\\-1.2}$ & 24.0$\substack{+1.2\\-1.2}$\\
        $\zeta_{\text{H}\alpha}$ & 0.35$\substack{+0.03\\-0.03}$ & 0.34$\substack{+0.03\\-0.03}$ & 0.29$\substack{+0.02\\-0.02}$\\
        $\zeta_{\text{H\,{\sc i}}}$ & 0.51$\substack{+0.02\\-0.03}$ & 0.49$\substack{+0.02\\-0.02}$ & 0.44$\substack{+0.02\\-0.02}$ \\
        \hline
    \end{tabular}
    \end{center}
\end{table}

 The derived radii ($r$) of the peaks of H$\alpha$ and H\,{\sc i} emission (as listed in Table \ref{tab:byproducts}) are 14.9$\substack{+0.4\\-0.2}$\,pc and 21.5$\substack{+1.7\\-1.2}$\,pc, respectively. Given that the overlap between atomic gas and H$\alpha$ emission is negligible, one may expect that the entire regions are eventually ionised by young massive stars, and that the relative sizes of H$\alpha$ and H\,{\sc i} emitting regions may be related to the contraction of the regions over time. Assuming little expansion following the onset of star formation, and that the atomic gas time-scale derived represents the majority of total time taken for star formation, we can estimate an average net rate of collapse over this time-scale of ${\rm d}r/{\rm d}t\sim 0.1$\,pc\,Myr$^{-1}$. This low rate of collapse is likely because atomic clouds are not gravitationally bound \citep{Kim2007} and do not form stars via a simple monolithic collapse, rather stars are formed across clouds as part of a scale-free hierarchical collapse of the ISM \citep[e.g.][]{Efremov1998,Ward2018,Ward2019}, while clouds may simultaneously continue to accrete material from the diffuse ISM.

 The peak concentration parameters (also known as filling factors) for the reference and target maps ($\zeta_{\text{ref}}$ and $\zeta_{\text{target}}$) are shown in the final two columns of Table \ref{tab:byproducts}. These concentration parameters are defined as:
 \begin{equation}
     \zeta = \frac{2r}{\lambda} \text{.}
 \end{equation}
 While we do not make direct use of these filling factors, the accurate measurement of $t_{\text{over}}$ requires that they fall below a certain maximum value derived in Appendix B of \citet{Kruijssen2018}. In this case, where $t_{\text{over}}/\tau = 0.01$, both tracers would require filling factors of $\zeta < 0.38$ (see Fig. B1 of \citealt{Kruijssen2018}) in order to limit the contamination of adjacent peaks to less than 5 per cent. This means that any measured value of $t_{\text{over}}$ is likely the result of regions that are overlapping spatially rather than in time and that $t_{\text{over}}$ must be taken as an upper limit, as found in the previous section.

\label{compos_Sct}

Using the parameters obtained with H{\sc eisenberg}, it is possible to derive a number of parameters using a combination of the parameters directly derived from this work and externally obtained constants.

For the unfiltered H\,{\sc i} column density map we measure an H\,{\sc i} gas mass of 4.4$\times$10$^{8}$\,M$_{\sun}$ within the region that H{\sc eisenberg} is applied, broadly consistent with the previously determined atomic gas masses of 4.8$\pm$0.2$\times$10$^{8}$\,M$_{\sun}$ and $4.41\pm0.09\times10^{8}$\,M$_{\sun}$ from \citet{Staveley-Smith2003} and \citet{Bruns2005}, respectively. From the filtered H\,{\sc i} map, we obtain a remaining compact gas mass of 4.8$\times$10$^{7}$\,M$_{\sun}$ before correcting for the filtering process. When the additional flux losses due to the filtering process (see Section~\ref{fourier}) are taken into account, this mass increases to $\sim$5.7$\times$10$^{7}$\,M$_{\sun}$. This implies that $\sim$10\% of the H\,{\sc i} mass of the LMC is contained within dense atomic clouds, with the remaining $\sim$87\% as diffuse emission. Given that the H\,{\sc i} gas mass makes up approximately 19\% of the total baryonic mass of the LMC (based on the H\,{\sc i} mass of \citet{Staveley-Smith2003} and the stellar mass of  \citet{Skibba2012}), this implies that compact (and potentially star forming) H\,{\sc i} clouds contain on the order of 2\% of all of the mass in the LMC. Optically thick H\,{\sc i} is also likely to be present in some of these regions. A recent study of H\,{\sc i} self-absorption in the Small Magellanic Cloud found a cold H\,{\sc i} fraction of $\sim$0.2 \citep{Jameson2019}, similar to those found in the Milky Way of $\sim$0.15--0.2 \citep{Heiles2003,Murray2015,Koley2019}. It is therefore possible that a similar fraction of optically thick atomic gas may be present in the LMC and remains unaccounted for here.

Similarly, a diffuse fraction of $\sim$86\% is measured for the H$\alpha$ emission map. However, because this map contains continuum emission as well as H$\alpha$ emission, a large portion of this apparent diffuse emission originates from the wider stellar population of the LMC. This is particularly evident within the LMC bar region in Fig. \ref{example_filtering}. In the left-hand image, where no filtering has been applied, the stellar component of the bar is clearly visible. However, in the right-hand image, after filtering in Fourier space, the stellar bar is suppressed significantly. This means that we cannot determine the exact diffuse fraction of H$\alpha$ emission using these data, but we can set an upper limit on this fraction of 87\%. While the diffuse H$\alpha$ emission cannot be directly linked to any single H\,{\sc ii} region, the majority of this emission is likely to originate from these star forming regions through photon leaking. This process is exacerbated in the low metallicity Magellanic Clouds due to the porous nature of the ISM in low metallicity galaxies \citep{Madden2006, Cormier2015, Dimaratos2015,Chevance2016}.

Using the atomic gas mass of \citet{Staveley-Smith2003}, $M_{\text{gas}}=4.8\pm0.2\times$10$^{8}$\,M$_{\sun}$ and the combined H$\alpha$ plus 24\,$\mu$m emission SFR of \citet{Whitney2008} (0.19\,$M_{\sun}$\,yr$^{-1}$) for stars younger than 10\,Myr, as well as the new empirically derived H\,{\sc i} cloud lifetime ($t_{\text{H{\sc i}}}=$48$\substack{+13\\-8}$\,Myr), it is possible to calculate a SFE per H\,{\sc i} cloud in the LMC, which is defined as ratio of the measured H\,{\sc i} cloud lifetime and the depletion time \citep{Kruijssen2018}:
\begin{equation}
    \epsilon_{\text{sf}} = \frac{\text{SFR}}{M_{\text{gas}}}t_{\text{H{\sc i}}} \text{,}
\end{equation}
where $M_{\text{gas}}$ is the total H\,{\sc i} gas mass, yielding an SFE of $\epsilon_{\text{sf}}=1.9\substack{+0.2\\-0.4}$\%. Assuming that only the measured compact H\,{\sc i} component (5.7$\times$10$^{7}$\,M$_{\sun}$) takes part in the star formation process yields an efficiency of 16$\substack{+5\\-3}$\%. Such an approach neglects the presence of molecular gas, cold atomic gas, and the continued accretion of material over the lifetime of a cloud. Therefore, these two values represent upper and lower limits on the SFE per H\,{\sc i} cloud of 16\% and 1.9\%, respectively.

\subsection{30 Doradus and the impact of localised, extreme star formation}
\label{30Dorsec}

30 Doradus is the brightest supergiant H\,{\sc ii} region in the Local Group of galaxies, with an ionising output that is equivalent to over a thousand O7V stars \citep{Crowther2019}.
Given the extreme nature of 30 Doradus, it is reasonable to expect that the region may have a significant impact on the average time-scale of H\,{\sc i} emission derived for the entire LMC. This could lead to the determination of a time-scale that is not relevant to either 30 Doradus or the remainder of the LMC \citep[e.g.][]{Kruijssen2018,Hygate2019inprep}. The technique employed in this paper is statistical in nature and therefore cannot be used to determine the cloud lifetime of a single region; however, we can mask the region surrounding 30 Doradus and therefore measure the impact of 30 Doradus on the derived average time-scale for the whole LMC.

\begin{figure*}
	\begin{minipage}{170mm}
		\begin{center}
	\includegraphics[width=0.57\linewidth]{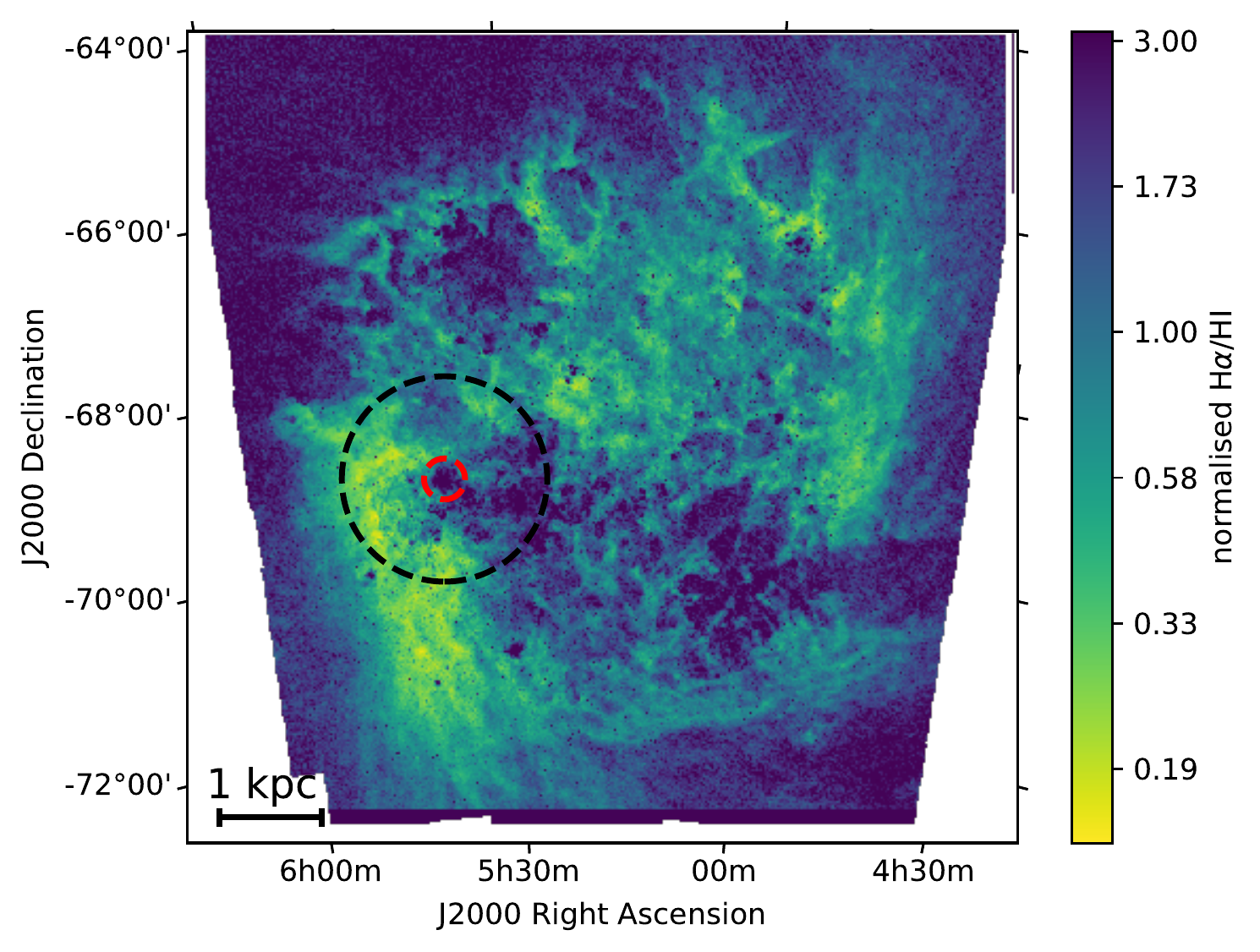}
	\includegraphics[width=0.42\linewidth]{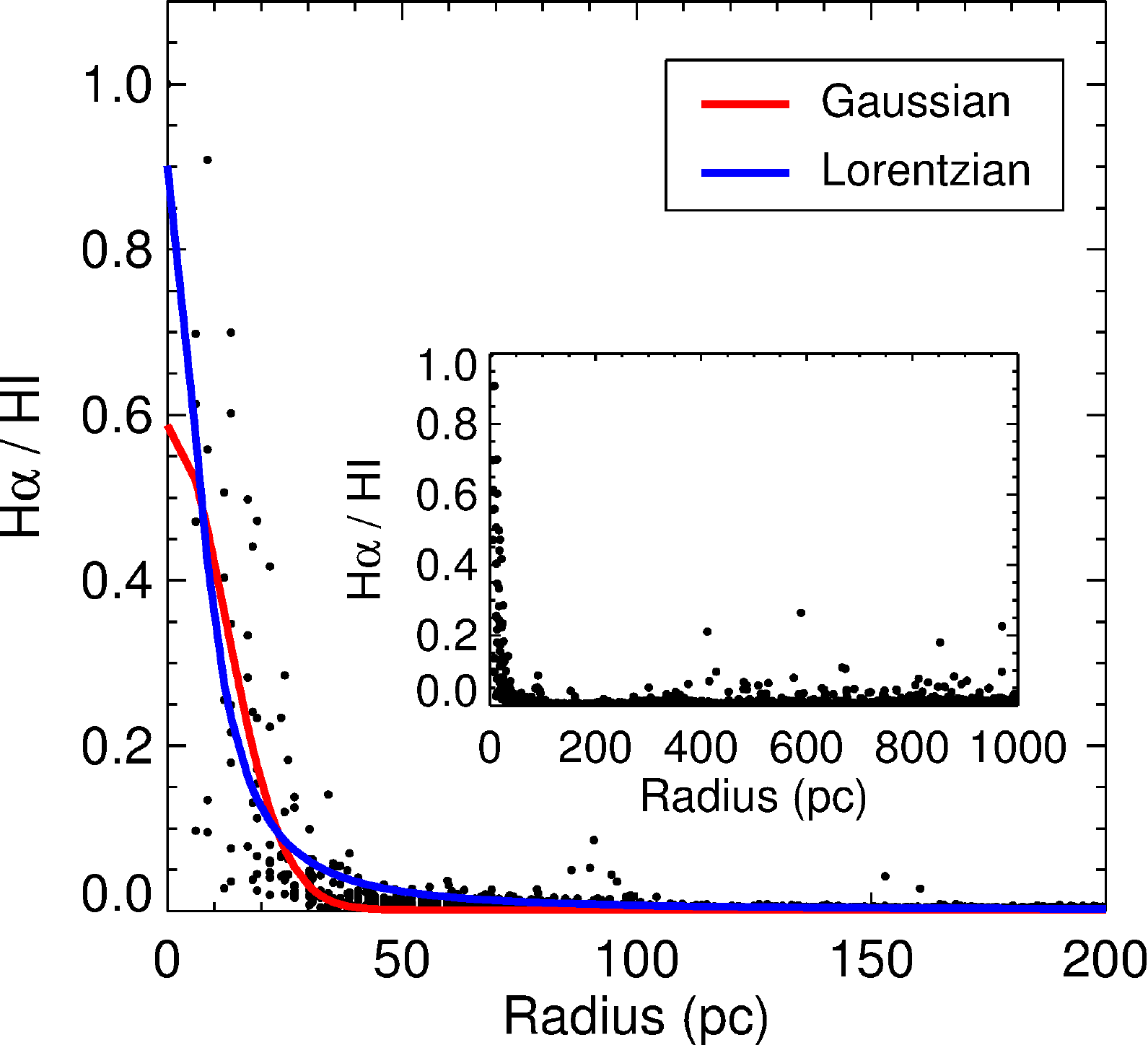}
	\caption{\label{HalphadivHI_fig} Left: the MCELS H$\alpha$ image of the LMC divided by the H\,{\sc i} image at a pixel scale of 50\arcsec ($\sim$12\,pc). Dark pixels indicate a high H$\alpha$-to-H\,{\sc i} flux ratio and light pixels indicate a low ratio. The black dashed circle marks the location of the 1\,kpc radius mask around 30 Doradus and the red dashed circle indicates the 200\,pc radius mask. Right: The radial profile of the H$\alpha$-to-H\,{\sc i} flux ratio within 200\,pc about the pixel in the 30 Doradus region that exhibits the highest H$\alpha$-to-H\,{\sc i} flux ratio, normalised by the peak ratio. Right, embedded panel: The radial profile of the H$\alpha$-to-H\,{\sc i} flux ratio within 1\,kpc about the pixel in the 30 Doradus region that exhibits the highest H$\alpha$-to-H\,{\sc i} flux ratio, normalised by the peak ratio. The Gaussian and Lorentzian functions fit to the radial profile of the H$\alpha$ / H\,{\sc i} emission ratio centred on 30 Doradus are shown in red and blue, respectively. Note that while the H$\alpha$-to-H\,{\sc i} ratio in 30 Doradus is among the highest in the LMC, the apparent size of 30 Doradus determined from this ratio is significantly smaller than the 800\,pc radius determined from ionisation parameter maps \citep{Pellegrini2012}.}
	\end{center}
	\end{minipage}
\end{figure*}

\begin{figure*}
\begin{minipage}{175mm}
  \includegraphics[width=0.5\linewidth]{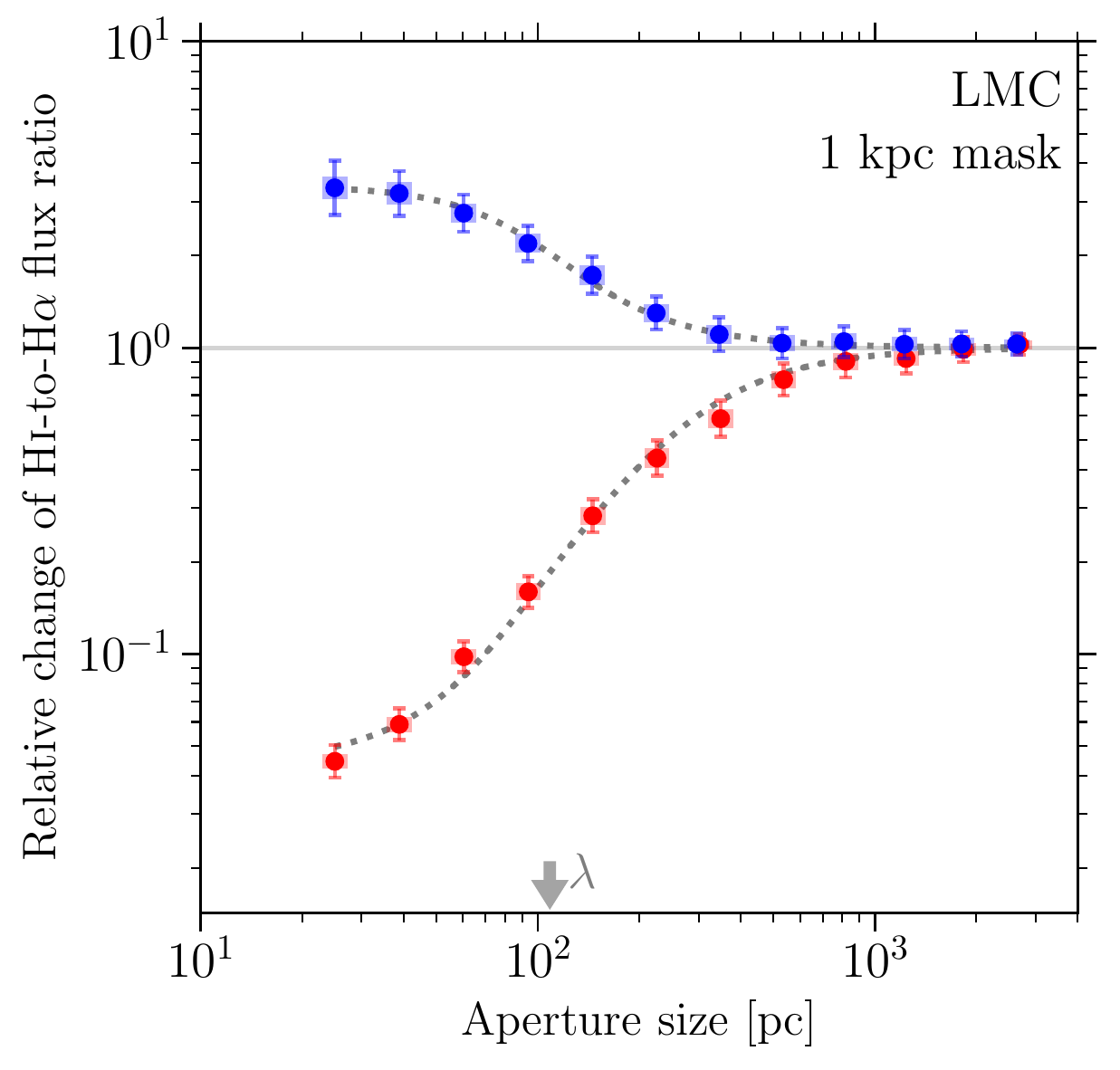}
  \includegraphics[width=0.5\linewidth]{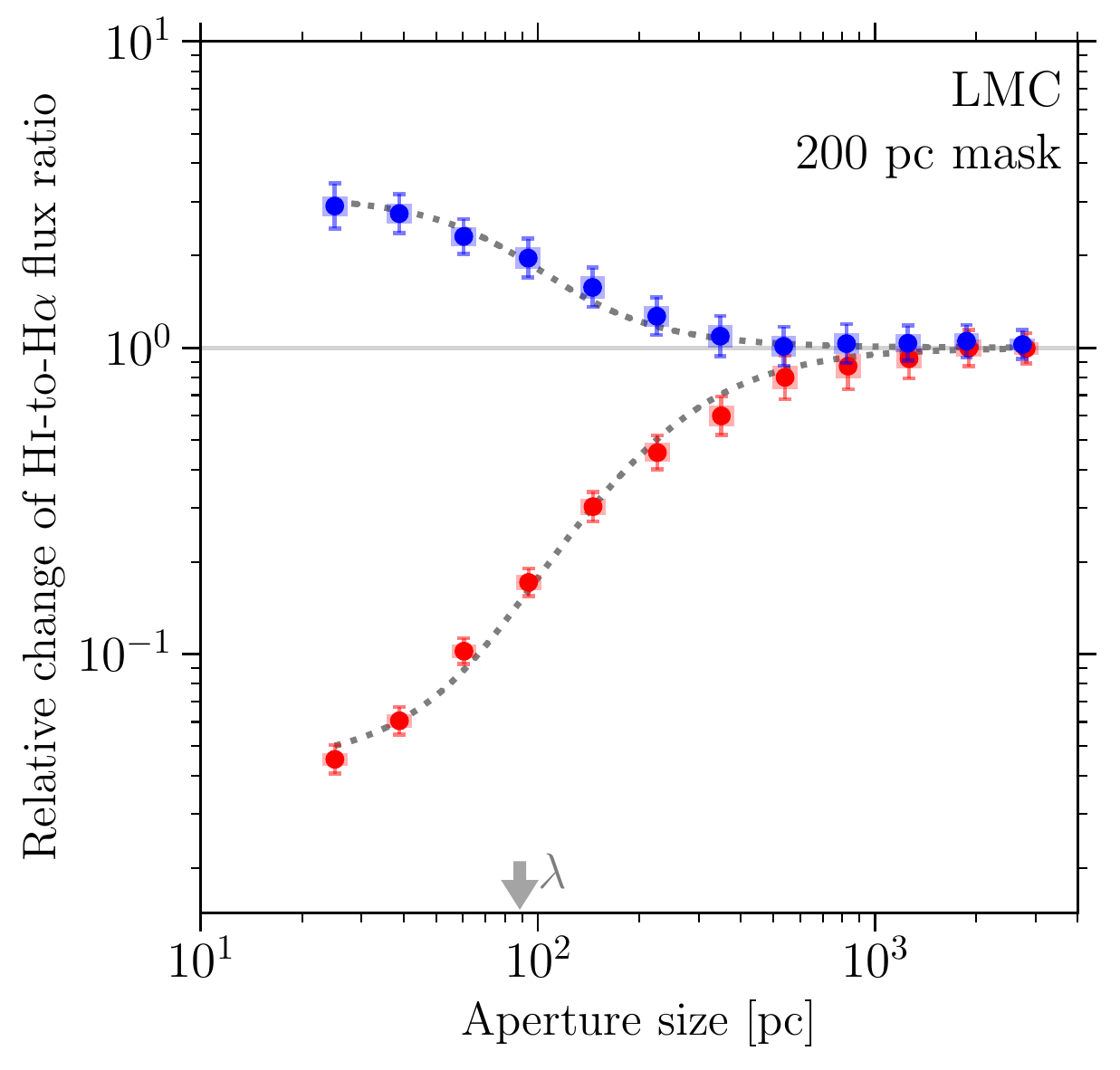}
  \caption{\label{Tuningfork_30dormask} Relative change of the gas-to-stellar flux ratio compared to the galactic average as a function of the aperture size when focusing on the H$\alpha$ peaks (red) and the H{\sc i} peaks (blue). The H{\sc eisenberg} tuning-fork-shaped model fit to the observed H\,{\sc i}-to-H$\alpha$ flux ratio versus aperture size is shown in green. Left: a 1\,kpc radius mask has been applied to remove the area surrounding 30 Doradus from the analysis. Right: a 200\,pc radius mask has been applied to remove the area surrounding 30 Doradus from the analysis. Note that the asymptote of the atomic gas branch falls higher when a 1\,kpc mask is used, indicating an increase in bias away from the galactic average when focusing on H\,{\sc i} peaks and therefore a shorter H\,{\sc i} emission time-scale is derived. The derived value of $\lambda$ is also longer when masking a 1\,kpc regions surrounding 30 Doradus.}
 \end{minipage}
\end{figure*}

\begin{table}
	\caption{ \label{30Dor_paramtab} The three fundamental fitted parameters from applying H{\sc eisenberg} to the LMC with 30 Doradus included, 30 Doradus removed using a 200\,pc mask, and 30 Doradus removed using a 1\,kpc mask. The second and third columns give the number of peaks selected in the H$\alpha$ and H\,{\sc i} maps, respectively. The derived parameters $t_{\text{H\,{\sc i}}}$, $t_{\text{over}}$, and $\lambda$ are given in columns 4, 5, and 6. The reduced $\chi^2$ value for each fit is listed in the final column. In each run, the H$\alpha$ plus continuum emission map is used as the reference map with an associated time-scale of 8.5$^{+1.0}_{-0.8}$\,Myr.}
	\begin{tabular}{l c c c c c c}
	\hline
	run & n$_{\text{H}\alpha}$ & n$_{\text{H\,{\sc i}}}$ & t$_{\text{H\,{\sc i}}}$ & t$_{\text{over}}$ & $\lambda$ & $\chi^{2}$ \\
		&	&	&	[Myr]	&	[Myr] &	[pc]	\\
	\hline
	No mask & 275 & 893 & 48$\substack{+13\\-8}$ & $<1.7$ & 86$\substack{+12\\-7}$ & 0.37 \\
	200\,pc mask& 274 & 834 & 50$\substack{+11\\-9}$ & $<1.9$ & 88$\substack{+11\\-8}$ & 1.09 \\ 
	1\,kpc mask & 217 & 737 & 38$\substack{+8\\-7}$ & $<2.1$ & 109$\substack{+11\\-10}$ & 0.84 \\
	\hline
	\end{tabular}
\end{table}

Through analysis of ionisation parameter maps based on the MCELS survey images of the LMC, \citet{Pellegrini2012} found that this ionising influence of the 30 Doradus star forming region extends up to 800\,pc from the central massive star cluster. To encompass this region, we have masked a region 1\,kpc in radius (shown in the left-hand panel of Fig. \ref{HalphadivHI_fig}), centred on 30 Doradus from both our H$\alpha$ and H\,{\sc i} emission maps. However, such a large mask also removes a large number of other emission peaks from the analysis, so we also run H{\sc eisenberg} using a 200\,pc (approximately double the derived characteristic separation length, $\lambda$) radius mask in order to remove only a single H$\alpha$ emitting peak from the analysis.

The resulting tuning fork-shaped model fits for the LMC with 30 Doradus masked (using 1\,kpc and 200\,pc radius masks) are shown in Fig \ref{Tuningfork_30dormask} using the MCELS H$\alpha$ image as the reference map. The resulting H\,{\sc i} time-scales, overlap time-scales, and separation lengths are given in Table \ref{30Dor_paramtab}.  When the 1\,kpc mask is utilised, we derive a significantly shorter time-scale. Additionally, a higher value of $\lambda$ is derived with the 1\,kpc mask. However, if only  H$\alpha$ emitting regions were removed from the analysis, one would expect that this would result in a longer derived H\,{\sc i} time-scale rather than a shorter one. Removing this 1\,kpc radius region from the analysis significantly reduces the derived H\,{\sc i} emission time-scale, but, it is likely that this is due to the removal of the large number of emission peaks, both H$\alpha$ emitting and H\,{\sc i}, that lie within the 1\,kpc radius of 30 Doradus. 

When using the 200\,pc mask, only a single H$\alpha$ emission peak is omitted from the analysis. As the large-scale ionising influence of 30 Doradus is largely due to the leakage of photons from the H\,{\sc ii} regions, such an approach is justified as larger-scale emission features have already been removed by the Fourier filtering discussed in Section \ref{fourier}. In this case the model fit is the best of all three runs (in terms of $\chi^{2}$) and all three of the fundamental free parameters ($t$,$t_{\text{over}}$, and $\lambda$) derived are wholly consistent with those derived when 30 Doradus is included in the analysis. At first glance this is somewhat surprising. Why is it that such an extreme star forming region has little or no effect on the measured average lifetime of atomic clouds in the LMC?

The left panel of Fig. \ref{HalphadivHI_fig} shows a map of H$\alpha$-to-H\,{\sc i} flux ratio for the LMC. When comparing this image to the MCELS H$\alpha$ mosaic of the LMC shown in the top-right panel of Fig.~\ref{example_filtering}, it is immediately clear that the extreme influence of 30 Doradus on the ratio between H$\alpha$ and H\,{\sc i} is limited to a relatively small area compared to the H$\alpha$ emission alone.
In the right panel of Fig. \ref{HalphadivHI_fig}, the normalised H$\alpha / $H\,{\sc i} flux ratio is shown against radial distance from the pixel with the highest H$\alpha / $H\,{\sc i} ratio in the 30 Doradus region. The impact of 30 Doradus on its immediate surroundings is extreme. The mean H$\alpha / $H\,{\sc i} ratio across the entire region of the LMC analysed with H{\sc eisenberg} is 0.2\% that of the 30 Doradus region. However, the effect of 30 Doradus on this ratio is negligible beyond a radius of $\sim$100\,pc. The radial profile of 30 Doradus has been fitted with a Gaussian distribution (red curve) and a Lorentzian distribution (blue curve), yielding profile dispersions of 12 and 8\,pc, respectively. We have therefore established that while the ionising influence of 30 Doradus reaches to extreme distances, the area over which it has a significant impact on the H\,{\sc i}-to-H$\alpha$ ratio is relatively small. As the derived characteristic length scale between independent regions is $\sim$100\,pc, 30 Doradus likely only affects the measured gas-to-stellar flux ratio within a single region (as removed by a 200\,pc mask; see Table~\ref{30Dor_paramtab}), $\sim$0.1\% of the total number of regions detected. While the effect of an extreme star forming region on the galaxy-wide gas-to-stellar flux ratio may be significant regardless of the number of regions within galaxies, it is clear that a single 30 Doradus-like region is unable to dominate the gas-to-stellar flux ratio across the entirety of the LMC.
This then explains the small effect of removing the immediate area surrounding 30 Doradus on the average time-scales calculated in this work. On the other hand, removing a larger area of 1\,kpc in radius has the effect of lowering the derived H\,{\sc i} cloud lifetime, most likely due to the removal from the analysis of 98 H\,{\sc i} clouds (as well as 75 H$\alpha$ emission peaks) in the vicinity of, but not coincident with, 30 Doradus. As Figures \ref{example_filtering} and \ref{HalphadivHI_fig} show, a large fraction of the H\,{\sc i} of the LMC originates from this region, which cumulatively outweighs the impact of removing a single extreme H$\alpha$-emitting region.

\subsection{Comparison with model time-scales}

\label{modelComparison}

Using the generalised theory for cloud lifetimes under the influence of galactic dynamics of \citet{Jeffreson2018}, we can estimate an expected lifetime based on the gas and stellar surface densities and the measured rotation curve of the LMC. While the theoretical framework of \citet{Jeffreson2018} is aimed at deriving the time-scales associated with molecular clouds, it does not distinguish between different phases of the ISM. In terms of the impact of galactic dynamics, it should be equally applicable to clouds of atomic gas, particularly in atomic gas-dominated environments like the LMC \citep{Roman-Duval2014}. In order to calculate the lifetimes predicted by the \citet{Jeffreson2018} model, we use the measured atomic gas surface density within the region to which we apply H{\sc eisenberg} (12.8\,M$_{\odot}$\,pc$^{-2}$), and we assume a stellar surface density of 58.2\,M$_{\odot}$\,pc$^{-2}$, based on the assumption that the vast majority of the stellar mass of the LMC resides within the 3.3\,kpc radius of this study and using the total stellar mass of 2$\times$10$^{9}$\,M$_{\odot}$ calculated by \citet{Skibba2012}. We also make use of the rotation curve derived by \citet{Alves2000} and take the average velocity dispersions to be 21\,km\,s$^{-1}$ \citep{Graff2000}, and 15.8\,km\,s$^{-1}$ \citep{Kim1998} for the stellar and gas components, respectively. We adopt a pattern speed of 40\,km\,s$^{-1}$kpc$^{-1}$ for a single spiral arm as derived by \citet{Dottori1996}. 

We calculate an average time-scale for the gravitational collapse of the mid-plane ISM of 55\,Myr for atomic gas in the LMC within the 3.3\,kpc region studied in the current work. Additionally, we predict time-scales of 203, 132, 885, and 84\,Myr for the effects of cloud-cloud collisions, epicyclic perturbations, spiral arm crossing, and shear, respectively, following \citet{Jeffreson2018}. Combining these time-scales as rates then yields a predicted average atomic cloud lifetime in the LMC of 50\,Myr. This is entirely consistent with our empirically derived atomic cloud lifetime of 48$\substack{+13\\-8}$\,Myr. Note that as shear acts to prevent the collapse of clouds, it is treated as a negative rate when combining timescales. The fact that the empirical measurement is in agreement with the theoretical cloud lifetime prediction, and that this time-scale is extremely close to the mid-plane ISM free-fall time-scale leads us to the conclusion that the lifetimes of atomic clouds in the LMC are primarily set by the local ISM free-fall time-scale and that the effects of cloud-cloud collisions, epicyclic perturbations, spiral arms, and shear are comparatively small because they are acting on much longer time-scales. 

\begin{figure*}
	\centering
	\includegraphics[width=0.8\linewidth]{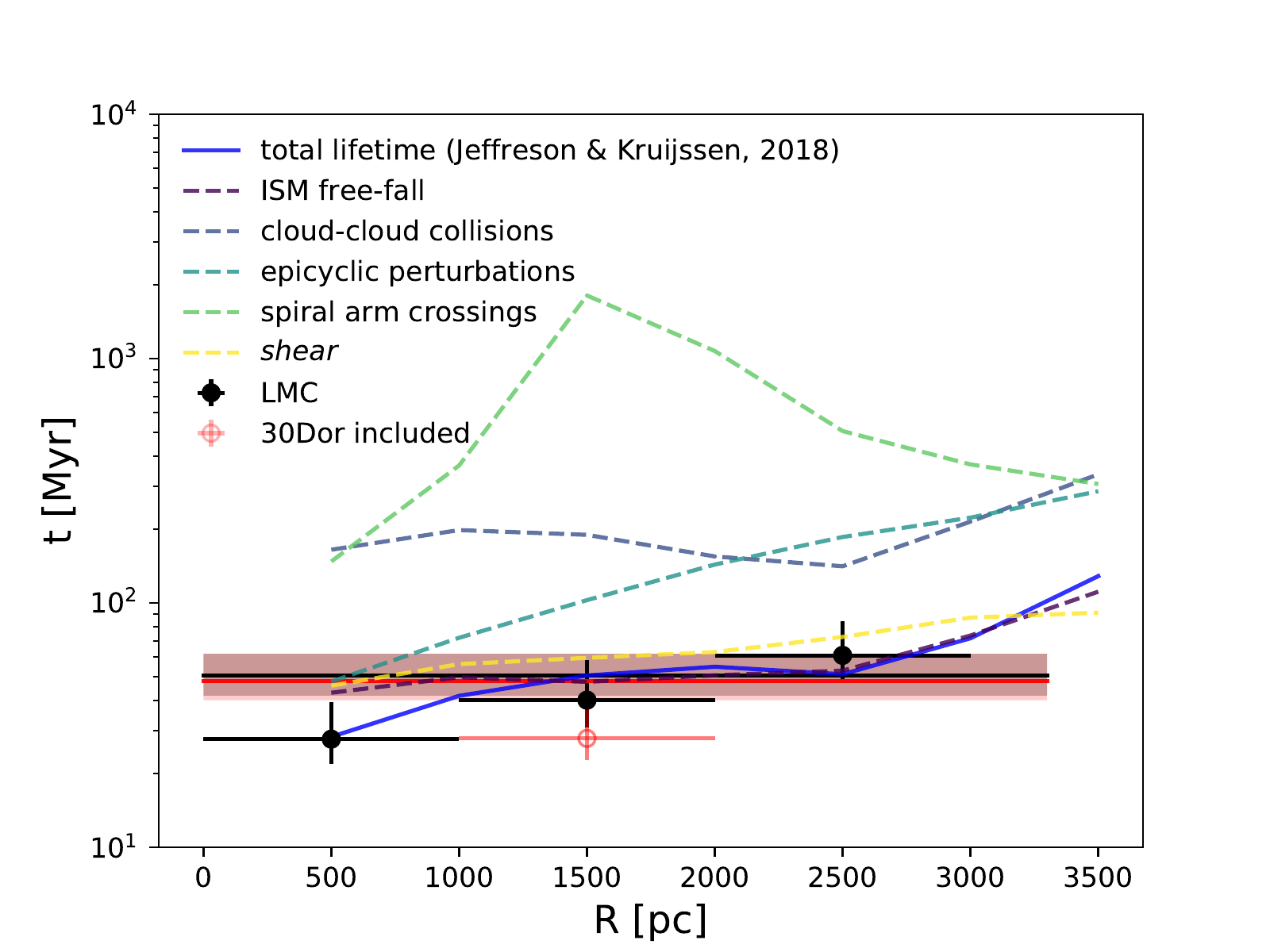}
	\caption{Predicted and measured atomic cloud lifetimes as a function of radius in the LMC. Dashed lines: predicted time-scales for the gravitational collapse of individual mechanisms (see the legend). The combined predicted atomic gas lifetime ($t$, \citealt{Jeffreson2018}) is shown in the solid blue line. The solid black and red lines show the measured average time-scale with and without excluding 30 Doradus, respectively, with 1$\sigma$ uncertainties denoted by the shaded region. Black points show the measured atomic cloud lifetimes for three 1000\,pc radial bins omitting 30 Doradus from the analysis and red points show the measured atomic cloud lifetimes with 30 Doradus.}
	\label{fig:theoreticaltscales}
\end{figure*}

The effects of various dynamic processes on the predicted time-scales associated with clouds are expected to vary significantly as a function of galactocentric radius. Therefore, we have divided the LMC into three radial bins of width 1000\,pc centred on radii of 500, 1500, and 2500\,pc and applied H{\sc eisenberg} to each bin individually. The resulting time-scales are shown in Fig. \ref{fig:theoreticaltscales} alongside the predicted time-scales calculated following \citet{Jeffreson2018}.

The atomic cloud time-scale is shortest in the innermost bin ($0<R<1000$\,pc),  likely due to the combination of  shorter epicyclic perturbations at small radii with the short free-fall time-scale. While the gravitational collapse time-scale is the shortest at all radii, in the inner-most bin only the combination of all time-scales successfully predicts the measured atomic cloud lifetime. In the intermediate bin ($1000<R<2000$\,pc), the theoretical time-scale over-predicts the cloud lifetime by more than 1$\sigma$ when including 30 Doradus. However, when the immediate region surrounding 30 Doradus is masked (using the 200\,pc mask), the predicted time-scale falls well within the uncertainties of the measured time-scale. While 30 Doradus has little impact on the galactic average, the effect of the region is far greater when it forms part of a limited radial range, resulting in a time-scale that differs by more than 1$\sigma$ depending on whether 30 Doradus is included. In the outermost radial bin ($2000<R<3000$\,pc), the predicted cloud lifetime is somewhat shorter than the measured cloud lifetime but is consistent considering the uncertainties on the observed value.

\section{Discussion}
\label{discussion}

In this Section, we first discuss the robustness of our results, i.e. to what extent they can be considered an accurate representation of atomic cloud lifetimes in the LMC. This is followed by a discussion of the meaning of these results within the wider context of the time-scales on which the star formation process takes place and, in particular, with respect to the estimates for molecular cloud lifetimes.

\subsection{Robustness of results}

\label{robostness}

In \citet{Kruijssen2018}, several criteria are outlined which should be met in order for the quantities derived by the H{\sc eisenberg} code to be considered robust. In this section we confirm that the results of this study comply with the subset of these criteria that is relevant in the context of our analysis. These criteria are listed here:

\begin{enumerate}
	\item The durations of the two phases in question must not differ by more than an order of magnitude. Larger differences lead to systematic biases in the derived time-scales. The time-scales in this work differ by a factor of $\sim$6. In contrast, the continuum subtracted H$\alpha$ emission time-scale differs from the derived H\,{\sc i} time-scale by an order of magnitude, and therefore using continuum subtracted H$\alpha$ emission as a reference map to determine the time-scale associated with H\,{\sc i} clouds may lead to a systematic bias in the results. In Appendix~\ref{Appendix_A}, we show that the results of using a continuum-subtracted H$\alpha$ map are consistent with those presented in Section \ref{results} but with increased uncertainties.
	
	\item The characteristic length scale $\lambda$ must satisfy $\lambda \geq N_{\text{res}}l_{\text{ap,min}} / \cos i$ 
	\,where $N_{\text{res}}$ is a number of resolution elements. $N_{\text{res}} = 1$ for $t$ or $N_{\text{res}} = 1.5$ for $t_{\text{over}}$. With values of $i < 30$\textdegree, $\lambda \sim 85$\,pc, and $l_{\text{ap,min}} = 25$\,pc, this criterion is easily met.
	
	\item The number of peaks must be equal to or greater than 15 to achieve an order-of-magnitude estimate of time-scales. Peak counts greater than or equal to 35 yield precisions better than 0.2\,dex. In this study, we always select more than 100 peaks for both the H$\alpha$ and H\,{\sc i} images, even when considering 1000\,pc radial bins. The actual precision is dependent on the number of peaks selected (with larger numbers of peaks yielding better precisions) and is included in the uncertainties given for the time-scales shown in Table \ref{Fun_tab}.
	
	\item The SFR as a function of age should not vary by more than $0.2$\,dex when averaged over age intervals with a width of $t_{\text{ref}}$ or $t$. This criterion ensures that the bias in the time-scales derived due to variations in SFR are less than 50\%. While the star formation history (SFH) is not something which we measure in this work (we only retrieve instantaneous measurements of current average time-scales), we can speculate as to whether this criterion is met by the LMC. The time-scales used and derived in this study are relatively short ($< 65$\,Myr) and therefore it is not expected that the SFH should vary significantly on these time-scales over an entire galaxy. \citet{Harris2009} determine the SFH of the LMC, both over the entire galaxy and for individual regions. Over the last 100\,Myr, the SFR of the LMC may have varied by a factor of 2, however, over the last 65\,Myr, the region that exhibits by far the largest change in SFR is 30 Doradus. As 30 Doradus does not significantly bias the derived H\,{\sc i} time-scale (see Section \ref{30Dorsec}), we conclude that it is unlikely that SFR variations significantly bias the results of this study and that any bias introduced likely falls well within the existing uncertainties in the determination of the H\,{\sc i} time-scale.
	
	\item Each independent region should emit in both tracers at some point in its lifecycle, but does not need to be identified as an emission peak. It is clear that all H$\alpha$ emitting regions must have once (before the assembly of molecular clouds and the formation of massive stars) been an overdensity of neutral hydrogen atoms. On the other hand, the reverse cannot be said -- it is entirely possible for overdensities of neutral hydrogen to never form massive stars. While individual emission peaks are not assumed to correspond to independent regions, on average the flux contained within apertures focused on these peaks should be representative of independent regions that satisfy this criterion. The effects of stochastic initial mass function (IMF) sampling are examined in detail in \citet{Haydon2018}, where it is shown that a mean star forming region mass of $M_{\text{r}}>1000$\,M$_{\sun}$ (Fig. 15 of \citealt{Haydon2018}) is sufficient for the effects of IMF sampling to be neglected in the calculation of a time-scale using the H{\sc eisenberg} code. In order to ensure that IMF sampling effects do not significantly impact the results presented here, both the star forming regions traced in H$\alpha$ emission and the H\,{\sc ii} regions that form from the selected atomic gas peaks must satisfy this criterion. Using equation 24 from \citet{Haydon2018}, we calculate a mean star forming region mass of 1400\,M$_{\sun}$, satisfying the above star forming region mass criterion to neglect IMF sampling. The mean H\,{\sc i} emission enclosed in the minimum aperture size around H\,{\sc i} peaks in the sample corresponds to a mass of 5.5$\times10^{4}$\,M$_{\sun}$. Assuming a SFE of 2\% (the lower limit derived in Section \ref{byproducts_etc}) yields a future star-forming region mass of 1090\,M$_{\sun}$, indicating that the mean future star-forming region mass should be sufficiently well sampled to ignore the effects of IMF sampling. Moreover, where molecular gas and/or optically-thick H\,{\sc i} is present, more stars will form than the observed atomic mass alone would suggest and star forming regions likely continue to accrete material from the diffuse ISM during the process of star formation. While the selected peaks do not necessarily correspond to independent star-forming regions, it is likely that the majority of the H\,{\sc i} emission peaks selected in this work will result in the production of H\,{\sc ii} regions that would be detectable as peaks in H$\alpha$ emission. Therefore, the measured atomic cloud time-scale is not significantly biased by the peak selection process.
	
	\item In order to obtain a value for $t_{\text{over}}$, rather than an upper (or lower) limit, we require $0.05 < t_{\text{over}} / \tau < 0.95$. This criterion is not met by the work presented in this paper as the value determined for $t_{\text{over}}$ is so low. We must therefore adopt an upper limit of the overlap time-scale of $t_{\text{over}}<1.7$\,Myr based on the results presented here.

\end{enumerate}

Having satisfied all of the above criteria for a robust application of our methodology (except for those relating to the overlap time-scale), we can conclude that the derived H\,{\sc i} time-scales presented in this work are suitably robust and that the upper limits placed on the overlap time-scales are the best possible given the available data.

\subsection{H\,{\sc i} time-scales in context}

We have determined that the mean time-scale associated with overdensities of atomic gas is 5.6$\substack{+1.4\\-0.8}$ times that of the time-scale associated with the MCELS H$\alpha$ emission map (including continuum emission). Using the characteristic time-scale for H$\alpha$ emission at the metallicity of the LMC derived by \citet{Haydon2018}, we quantify this time-scale for H\,{\sc i} emission as 48$\substack{+13\\-8}$\,Myr. 

Interpreting the above time-scale requires some consideration of the virial state of the atomic clouds that make up the sample. \citet{Kim2007} found that the majority of H\,{\sc i} clouds in the LMC are supervirial. Similarly, if we adopt the average velocity dispersion determined by \citet{Kim2007} of 2.3\,km\,s$^{-1}$ as well as our average cloud mass and radius (24400\,\,M$_{\sun}$ and 21.5\,pc, respectively), we calculate a virial parameter of 5.6. {The H\,{\sc i} clouds presented in this work therefore appear to be supervirial. However, such a simplistic approach neglects the acceleration due to the molecular phase of the ISM and the stellar disc which is particularly important in low-density environments (e.g. \citealt{Schruba2019}).} The consistency between the measured time-scale associated with the atomic clouds and the predicted time-scale for the gravitational collapse of the ISM (the ISM free-fall timescale in Fig.~\ref{fig:theoreticaltscales}) following the approach of \citet{Jeffreson2018} (as shown in Section \ref{modelComparison}) suggests that while they are not simple isolated virialised systems, the clouds identified in this work are significant overdensities of a continuous hierarchical interstellar medium that represent the most likely sites of future star formation.

The lifetime of overdensities of atomic gas must necessarily provide an upper limit on the  formation time-scale of molecular clouds. Our derived time-scale for atomic gas of 48$\substack{+13\\-8}$\,Myr is certainly consistent with predictions for the formation time-scales of molecular clouds of tens of Myr \citep[e.g.][]{Larson1994,Goldsmith2007} as well as estimates of the molecular cloud lifetime in the LMC by \citet{Kawamura2009}. In future work, combining our measured atomic cloud lifetime with a similarly determined overlap time-scale for molecular clouds and atomic clouds in the LMC will provide the first empirically-determined molecular cloud formation time-scale.

Based on the NANTEN survey of molecular clouds in the LMC \citep{Fukui2008}, \citet{Kawamura2009} estimate a molecular cloud lifetime of 20${-}$30\,Myr. While we defer a measurement of the time-scale for which atomic and molecular clouds overlap to a follow-up study, we can use this estimate to place limits on the molecular cloud formation time-scale. Assuming no overlap time, implying an instant phase change from atomic to molecular gas, we use our result of $t_{\text{H{\sc i}}}=48\substack{+13\\-8}$\,Myr to place an upper limit on the molecular cloud formation time-scale of $\sim$50\,Myr from the onset of atomic clouds to the emergence of isolated molecular clouds. Assuming that atomic gas is present for the entirety of the molecular cloud phase and using the upper estimate of \citet{Kawamura2009}, we place a lower limit of $\sim$10\,Myr on the time-scale for cloud condensation prior to the emergence of CO emitting molecular clouds. While very loose constraints, these observationally derived limits are consistent with  predictions of molecular cloud formation time-scales (10${-}$50\,Myr; \citealt{Larson1994,Goldsmith2007}).

In a future paper, we will use H{\sc eisenberg} to empirically derive a time-scale for CO emitting molecular clouds in the LMC. This, combined with the time-scales derived in this work, will allow for much stronger constraints to be placed on the molecular cloud assembly time. This analysis will also constrain the time-scale over which molecular clouds are destroyed following the formation of massive stars. 

\section{Conclusions}
\label{conclusions}

In this paper, we have applied the uncertainty principle for star formation, originally presented in \citet{KL14}, using the H\,{\sc eisenberg} code \citep{Kruijssen2018}, to determine the average lifetime for atomic H\,{\sc i} clouds in the LMC, along with the time for which the neutral H\,{\sc i} emission overlaps with the H$\alpha$ emitting phase, and the characteristic separation length between sites of star formation in the LMC. Through masking certain regions from the input images, we have investigated the effects of the extreme star forming region 30 Doradus on these average values. Our key results are summarised below.

\begin{enumerate}
	\item We determine a lifetime of overdensities of atomic gas in the LMC that is 5.6$\substack{+1.4\\-0.8}$ times that of the time-scale associated with the reference H$\alpha$ plus continuum emission map from MCELS. Using the time-scale for H$\alpha$ (including continuum) by \citet{Haydon2018} yields an H\,{\sc i} time-scale of 48$\substack{+13\\-8}$\,Myr. This time-scale is entirely consistent with the predicted cloud lifetime following the approach of \citet{Jeffreson2018} that is dominated by the time-scale for the gravitational collapse of the mid-plane ISM.
	
	\item Combining the above H\,{\sc i} time-scale with previous estimates of the time-scale associated with molecular gas clouds in the LMC yields a formation time-scale for molecular clouds in the range 10${-}$50\,Myr.
	
	\item Using our derived H\,{\sc i} timescale (48$\substack{+13\\-8}$\,Myr), we determine that the SFE per H\,{\sc i} cloud in the LMC is of the order of 1.9$\substack{+0.2\\-0.4}$ per cent.
	
	\item The overlap time-scale for which H$\alpha$ emission and H\,{\sc i} emission co-exist is short with an upper limit of $<$1.7\,Myr. This is indicative of a near-complete phase change from neutral atomic gas into molecular and ionised gas by the onset of the H\,{\sc ii} region phase. Given that current estimates place the molecular cloud time-scale at $>$10\,Myr \citep[e.g.][]{Kawamura2009,Meidt2015,Kruijssen2019,Chevance2019}, this is an indication that there is an isolated molecular cloud phase that is not associated with atomic gas peaks prior to the onset of H\,{\sc ii} regions.
	
	\item The characteristic separation length between independent star forming regions in the LMC is $\lambda = 86\substack{+12\\-7}$\,pc. This is consistent with the previously observed break in the power spectrum of H\,{\sc i} in the LMC \citep{Elmegreen2000} and may be linked to the scale height of the disc.
	
	\item The effect of 30 Doradus on the H\,{\sc i} emission time-scale derived in this study is small. The region also has little effect on either the overlap time-scale or the characteristic length scale $\lambda$. A map of the H$\alpha$-to-H\,{\sc i} flux ratio for the LMC shows that, while the influence of 30 Doradus on the ionisation properties of the gas has been shown to extend to a radius of at least 800\,pc, the effect on the ratio between H\,{\sc i} and H$\alpha$ emission extends to a radius of less than 100\,pc. Despite the extreme ionising influence of 30 Doradus, as a single bright H\,{\sc ii} region, it does not dominate the measured average cloud lifetime when the entire LMC is consisdered.
	When a smaller number of regions are considered (as is the case when splitting the LMC into radial bins), the effects of 30 Doradus on the derived average time-scale does become significant at the 1$\sigma$ level.
\end{enumerate}

The work presented in this paper represents the first step towards a comprehensive, multi-tracer timeline of star formation. By empirically determining the average time-scale for atomic gas emission peaks in the LMC, we have begun populating this new timeline with the very earliest stage of the star formation process.
In subsequent papers we will investigate the application of the same technique to a number of tracers of different stages of the star formation process over a range of wavelengths, gradually incorporating more time-scales and subsequently derived parameters into our framework, culminating in the establishment of a robust multi-tracer timeline of star formation and feedback in the LMC.

\section*{Acknowledgements}

JLW and JMDK gratefully acknowledge support from the Deutsche Forschungsgemeinschaft (DFG, German Research Foundation) -- Project-ID 138713538 -- SFB 881 (\textquotedblleft The Milky Way System\textquotedblright, subproject B2).
MC and JMDK gratefully acknowledge funding from the DFG through an Emmy Noether Research Group (grant number KR4801/1-1) and the DFG Sachbeihilfe (grant number KR4801/2-1). JMDK and APSH gratefully acknowledge funding from the European Research Council (ERC) under the European Union's Horizon 2020 research and innovation programme via the ERC Starting Grant MUSTANG (grant agreement number 714907).
APSH is a fellow of the International Max Planck Research
School for Astronomy and Cosmic Physics at the University
of Heidelberg (IMPRS-HD).

\section*{Data availability}

No new data were generated in support of this research.

%%%%%%%%%%%%%%%%%%%%%%%%%%%%%%%%%%%%%%%%%%%%%%%%%%

%%%%%%%%%%%%%%%%%%%% REFERENCES %%%%%%%%%%%%%%%%%%

% The best way to enter references is to use BibTeX:

\bibliographystyle{mnras}
\bibliography{bibliography} % if your bibtex file is called example.bib

% Alternatively you could enter them by hand, like this:
% This method is tedious and prone to error if you have lots of references
%\begin{thebibliography}{99}
%\bibitem[\protect\citeauthoryear{Author}{2012}]{Author2012}
%Author A.~N., 2013, Journal of Improbable Astronomy, 1, 1
%\bibitem[\protect\citeauthoryear{Others}{2013}]{Others2013}
%Others S., 2012, Journal of Interesting Stuff, 17, 198
%\end{thebibliography}

%%%%%%%%%%%%%%%%%%%%%%%%%%%%%%%%%%%%%%%%%%%%%%%%%%

%%%%%%%%%%%%%%%%% APPENDICES %%%%%%%%%%%%%%%%%%%%%

\appendix

\section{Consistency of reference time-scales}

\label{Appendix_A}

In this appendix, we test the consistency of the H$\alpha$ reference time-scales determined using the relations of \citet{Haydon2018} for continuum subtracted images and those including continuum emission. This allows us to confirm that the longer timescales associated with H$\alpha$ emission maps that include continuum emission predicted by \citet{Haydon2018} are consistent with observations of the LMC. We also use the continuum subtracted SHASSA H$\alpha$ emission map to independently determine the time-scale associated with H\,{\sc i} emission in the LMC in order to test consistency between all three images used in this work. Finally, we use the H\,{\sc i} map to re-derive the timescale associated with the MCELS H$\alpha$ map in order to confirm that our methodology is commutative.

The SHASSA H$\alpha$ map has undergone continuum subtraction, while the MCELS H$\alpha$ image has not. In the case where  two tracers are always overlapping (as in the case of two H$\alpha$ emission maps), H{\sc eisenberg} is unable to provide an accurate fit, as one of the two branches either exhibits no bias with small aperture sizes, or may even exhibit a negative bias in the case that the peak sizes of one map are larger than the those in the other.
In order to test the consistency of the time-scales between these two maps, we take advantage of the non-degeneracy of the three fundamental parameters derived by the H{\sc eisenberg} code. As we only wish to test the consistency of the two time-scales, we rotate the SHASSA image by a 90$^\circ$ angle (where the final result does not strongly depend on the angle chosen). While we lose any information regarding the overlap time-scale ($t_{\text{over}}$) during which both tracers co-exist and the separation length ($\lambda$), this allows the relative time-scales between the two phases to be accurately constrained. This allows us to determine the relative time-scales between the continuum subtracted H$\alpha$ image and the H$\alpha$ image with continuum emission included. 

\begin{figure*}
\includegraphics[width=0.49\linewidth]{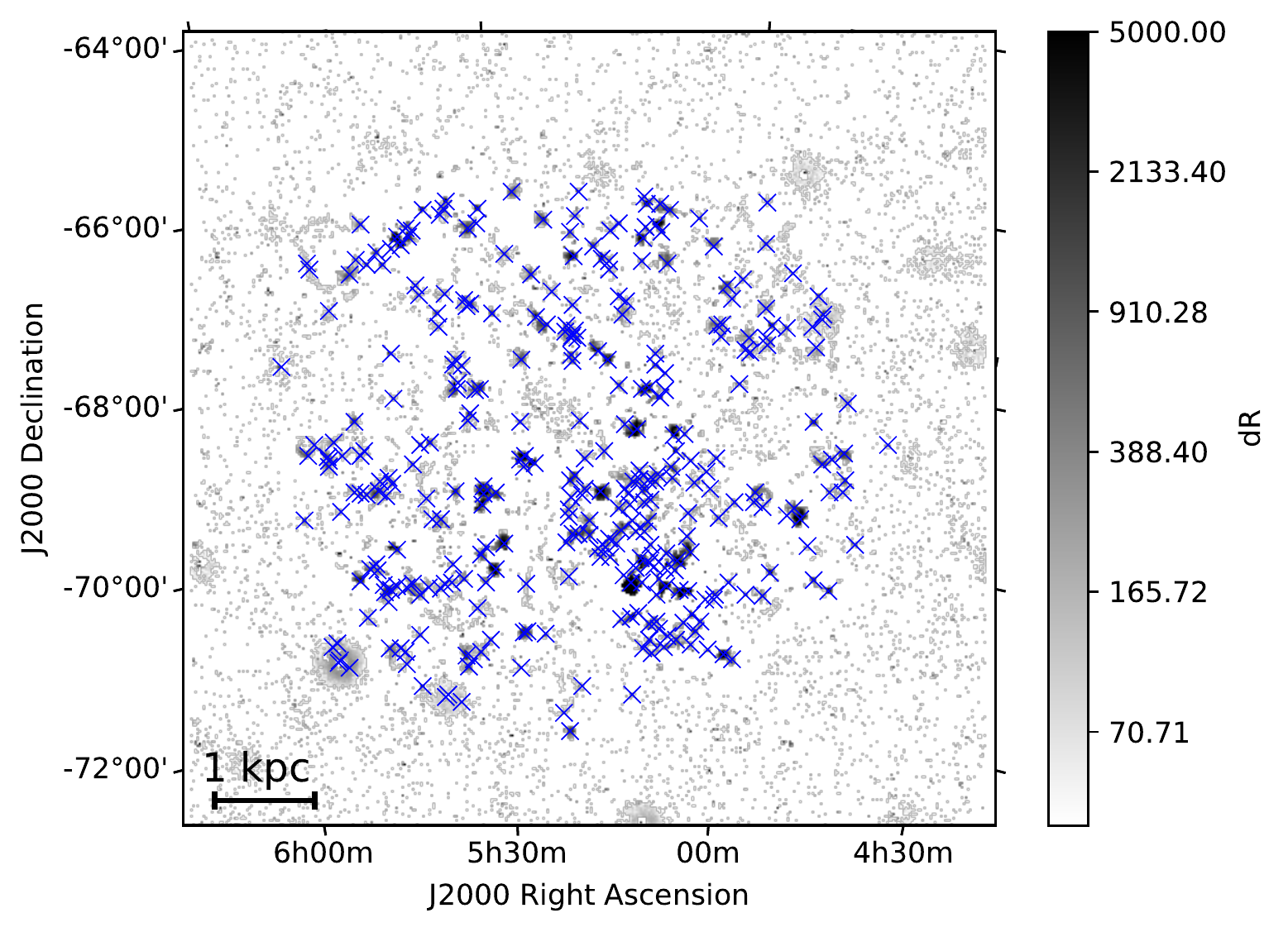}
\includegraphics[width=0.49\linewidth]{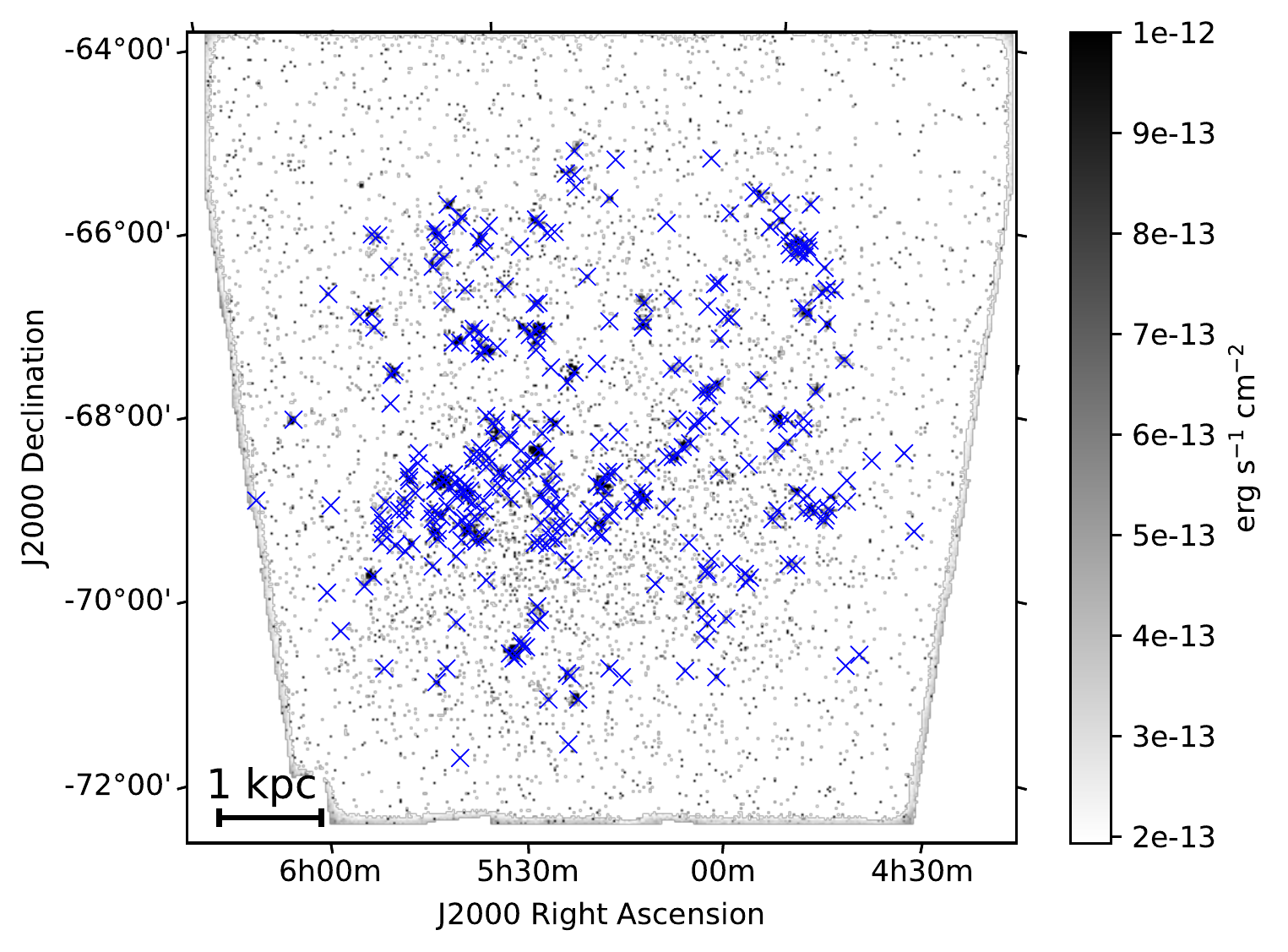}
\includegraphics[width=0.49\linewidth]{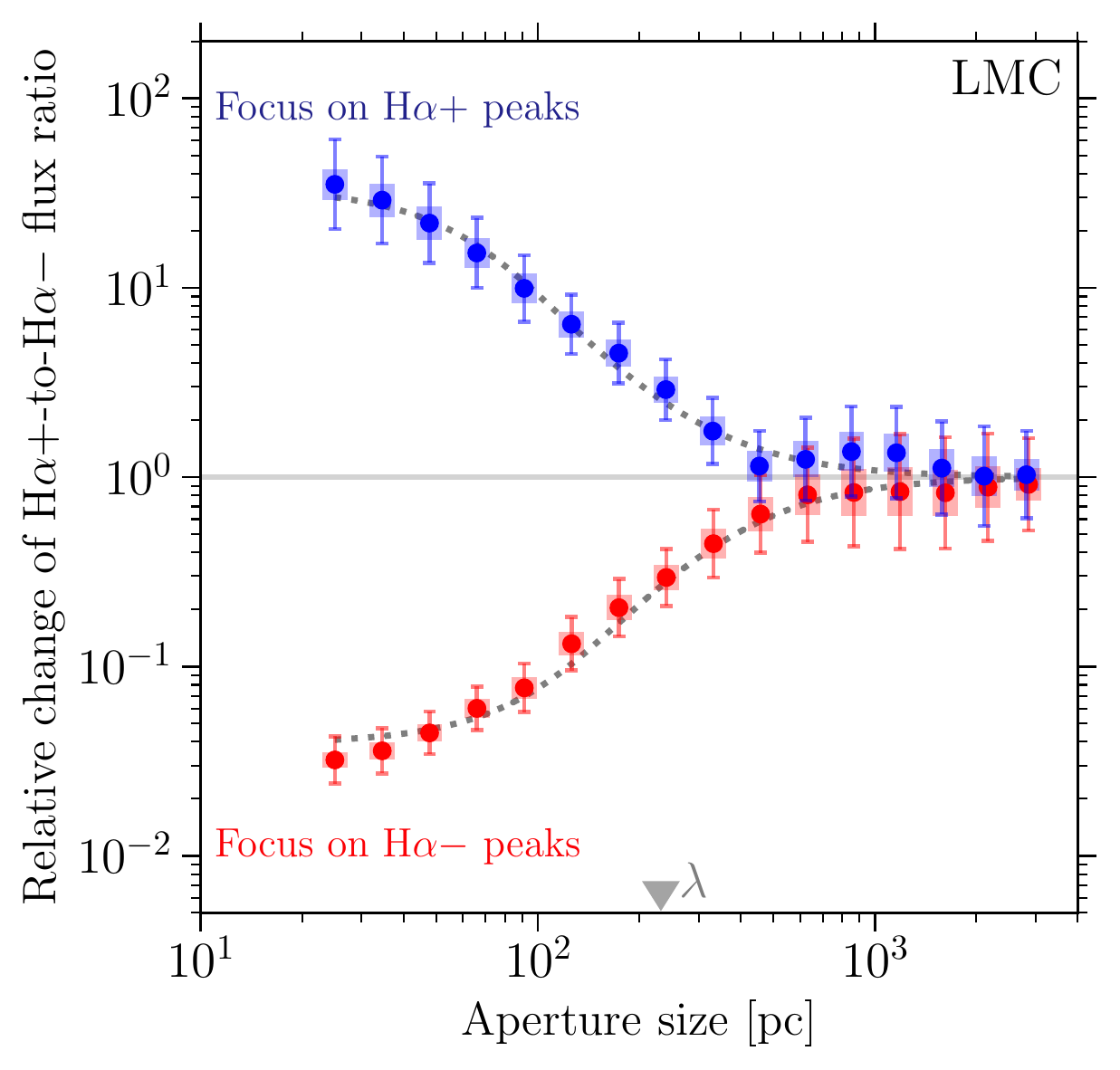}
\includegraphics[width=0.49\linewidth]{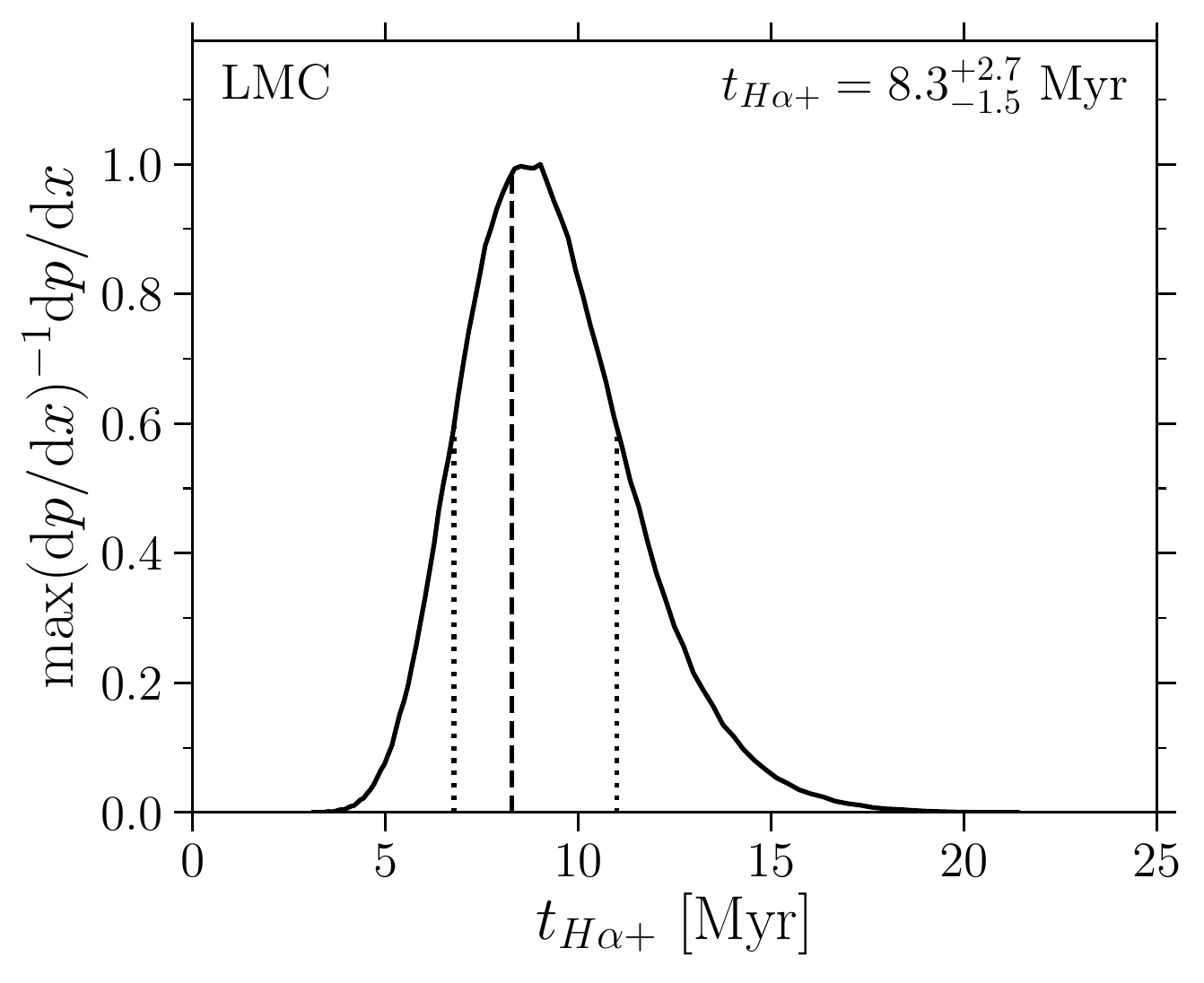}
\caption{\label{SHASSA_vs_MCELS} Top left: SHASSA continuum-subtracted H$\alpha$ image at 25\,pc resolution, rotated by 90\textdegree, with selected peaks marked with blue crosses. Top right: MCELS H$\alpha$ image at 25\,pc resolution with selected peaks marked with blue crosses. Bottom left: As in Fig. \ref{Tuningfork}, using the SHASSA H$\alpha$ image as a reference map (H$\alpha -$) and the MCELS H$\alpha$ plus continuum image (H$\alpha +$) as a target map. Bottom right: The PDF for the derived value of $t_{\text{H}\alpha +}$.}
\end{figure*}

In Fig. \ref{SHASSA_vs_MCELS}, we show the results of the H{\sc eisenberg} run using the SHASSA continuum-subtracted H$\alpha$ image rotated by 90\textdegree\, as the reference map to determine the time-scale associated with the MCELS H$\alpha$ map of the LMC. The reference time-scale used in this run is 4.67$\substack{+0.15\\-0.34}$\,Myr, as determined using eqn. 11 of \citet{Haydon2018}. We determine a time-scale for H$\alpha$ emission plus continuum emission of 8.3$\substack{+2.7\\-1.5}$\,Myr. This is consistent with the 8.5$\substack{+1.0\\-0.8}$\,Myr time-scale determined for the MCELS H$\alpha$ filter using eqn. 12 from \citet{Haydon2018} within the derived uncertainties. This means that the observed ratio of the H$\alpha$ emission time-scales with and without continuum subtraction (0.57$\substack{+0.15\\-0.12}$) is fully consistent with that predicted by \citet[0.55$\substack{+0.08\\-0.08}$,][]{Haydon2018}.

\begin{table*}
\caption{The fundamental parameters derived from the tests performed in this appendix. The images used for the reference time-scale and the derived time-scale are listed in the first and second columns, respectively. H$\alpha+$ refers to the MCELS H$\alpha$ mosaic, and H$\alpha-$ refers to the continuum subtracted SHASSA H$\alpha$ image. The third and fourth columns list the number of peaks selected in each map, followed by the reference time-scales in the fifth column. The following three columns contain the fundamental parameters $t$, $t_{\text{over}}$, and $\lambda$. The reduced $\chi ^{2}$ values are listed in the final column.}
    \label{tab:consistenttable}
\centering
    \begin{tabular}{l c c c c c c c c}
    \hline
         ref & target & $n_{\text{ref}}$ & $n_{\text{target}}$ & $t_{\text{ref}}$\,[Myr] & $t$\,[Myr] & $t_{\text{over}}$\,[Myr] & $\lambda$\,[pc] & $\chi^2$ \\
         \hline
         H$\alpha -$ & H$\alpha +$ & 330 & 312 & 4.67$\substack{+0.15\\-0.34}$ & 8.3$\substack{+2.7\\-1.5}$ & 0.01$\substack{+0.14\\-0.01}$ & 232$\substack{+20\\-18}$ & 0.66 \\
         H$\alpha -$ & H\,{\sc i} & 319 & 951 & 4.67$\substack{+0.15\\-0.34}$ & 47.2$\substack{+30.6\\-10.7}$ & 1.2$\substack{+2.0\\-0.6}$ & 76$\substack{+16\\-9}$ & 0.22 \\
         H\,{\sc i} & H$\alpha +$ & 892 & 275 & 47.8$\substack{+13.2\\-8.0}$ & 8.6$\substack{+2.7\\-2.2}$ & 0.6$\substack{+0.8\\-0.4}$ & 87$\substack{+13\\-6}$ & 0.37 \\
         \hline
    \end{tabular}
\end{table*}

\begin{figure*}
\includegraphics[width=0.49\linewidth]{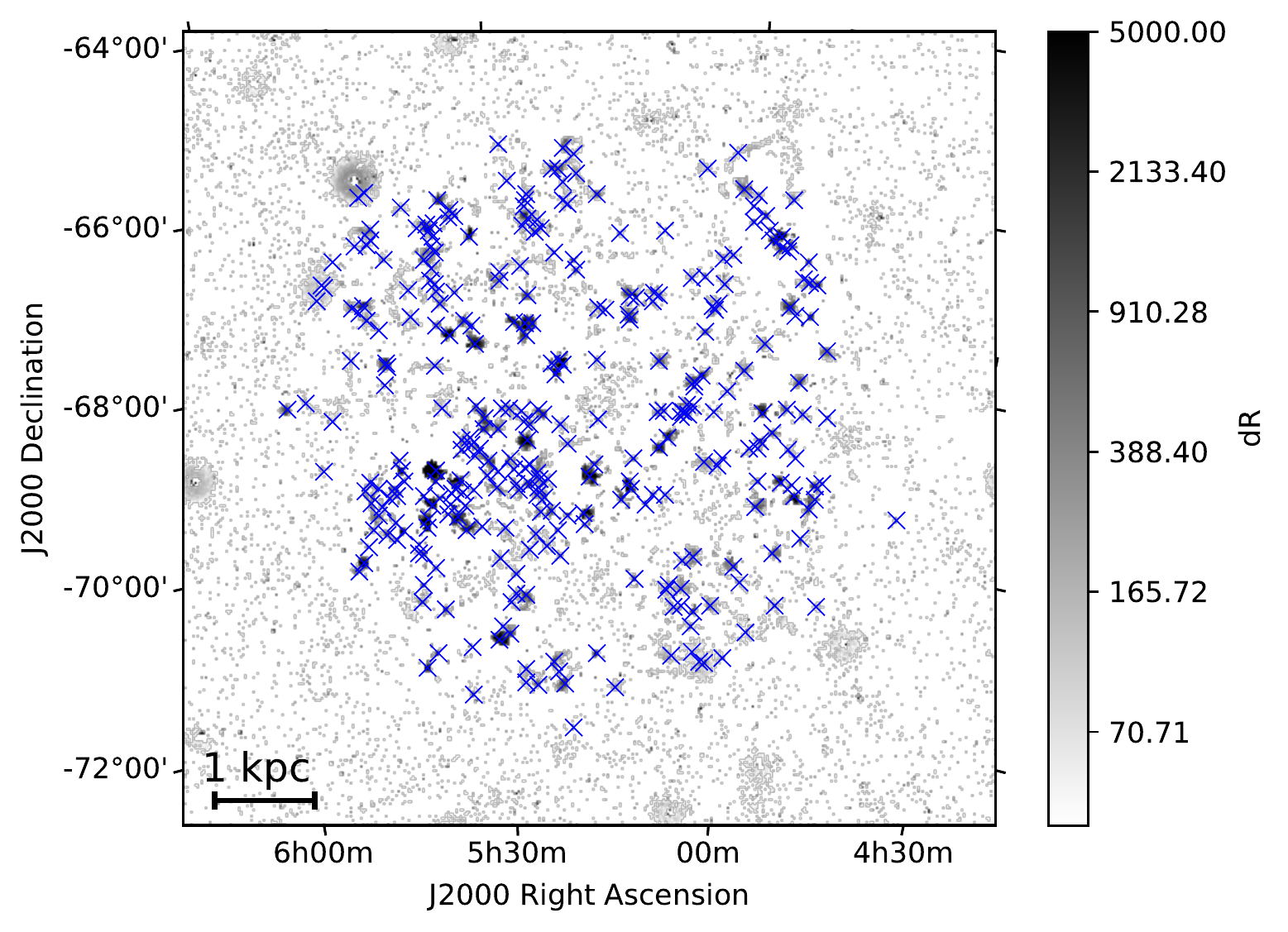}
\includegraphics[width=0.49\linewidth]{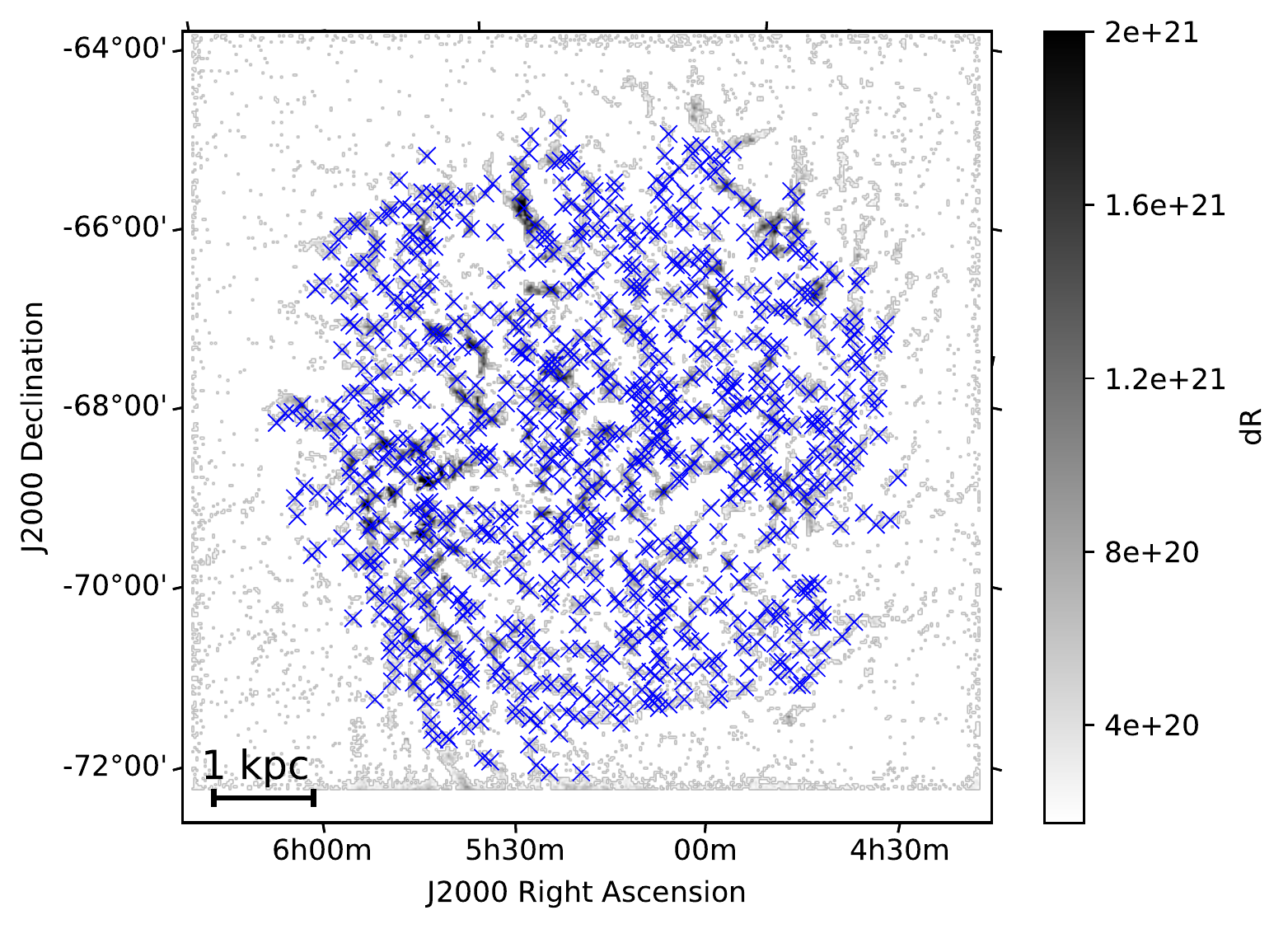}
\includegraphics[width=0.49\linewidth]{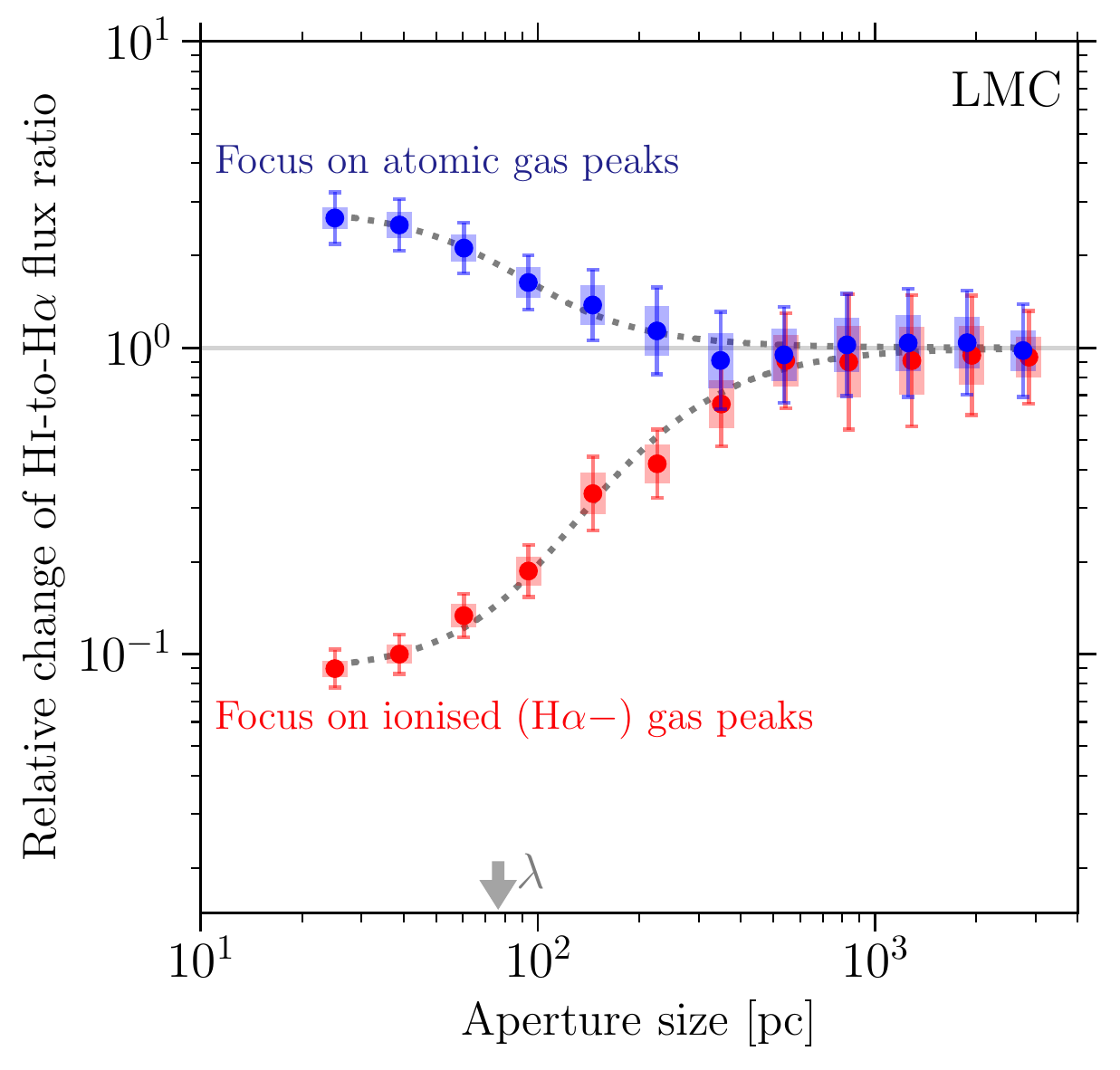}
\includegraphics[width=0.49\linewidth]{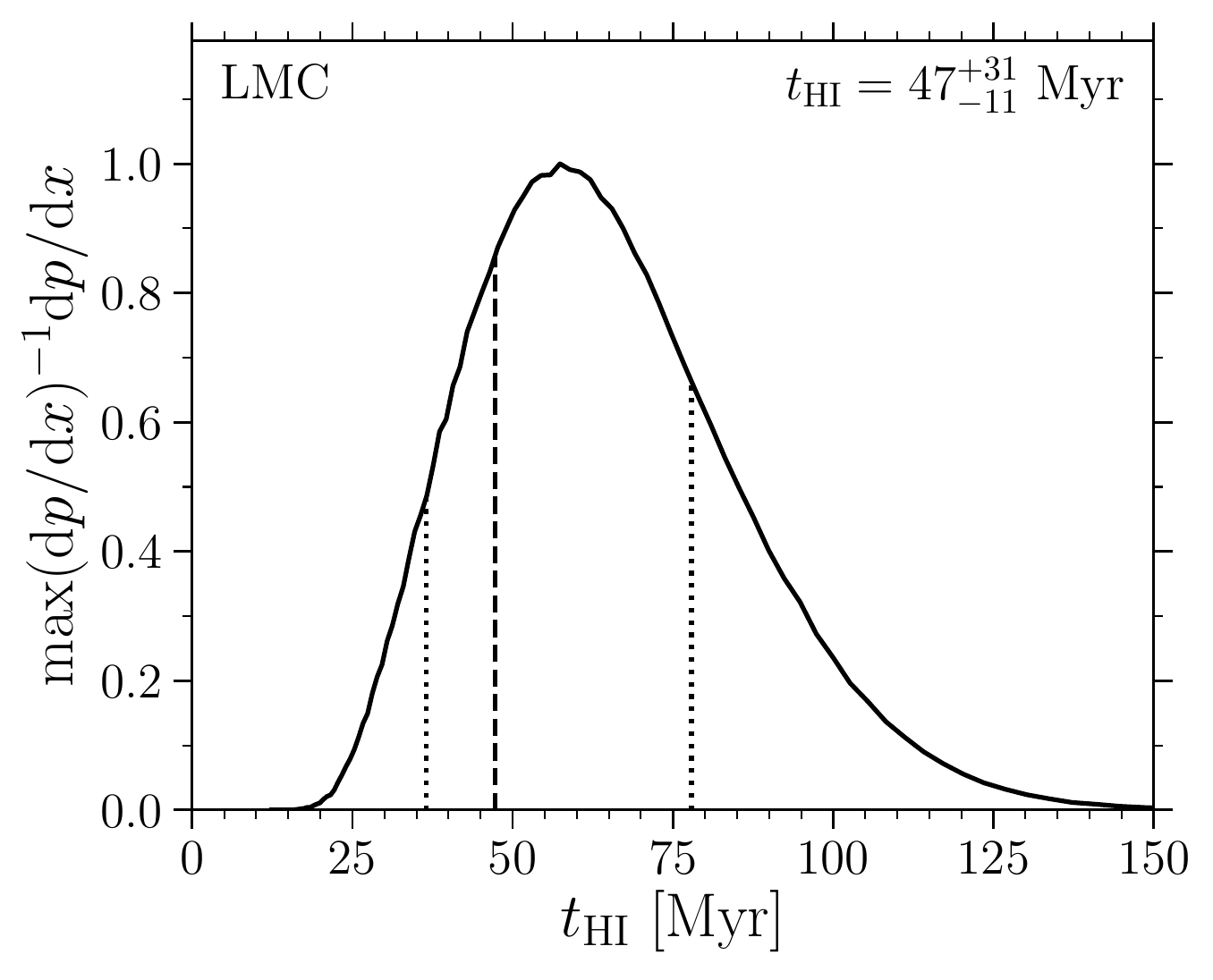}
\caption{\label{SHASSA_vs_HI} Top left: SHASSA H$\alpha$ image at 25\,pc resolution with selected peaks marked with blue crosses. Top right: H\,{\sc i} image at 25\,pc resolution with selected peaks marked with blue crosses. Bottom left: As in Fig. \ref{Tuningfork}, using the SHASSA H$\alpha$ image as a reference map and H\,{\sc i} as a target map. Bottom right: The PDF for the derived value of $t_{\text{H\,{\sc i}}}$.}
\end{figure*}

In Fig. \ref{SHASSA_vs_HI}, we show the results of the H{\sc eisenberg} run using the SHASSA continuum subtracted H$\alpha$ image to determine the characteristic time-scale for H\,{\sc i} emission peaks, with a reference time-scale of 4.67$\substack{+0.15\\-0.34}$~Myr. In this case, an H~{\sc i} cloud lifetime of 47$\substack{+31\\-11}$\,Myr is determined with a corresponding overlap time-scale of $t_{\text{over}}<3.2$\,Myr. While the uncertainties are significantly larger, these results are entirely consistent with those presented in Section \ref{results}. However, it should be noted that due to the relatively short time-scale of continuum subtracted H$\alpha$ emission compared to the derived H\,{\sc i} time-scale, this run does not satisfy the first of the criteria in Section 5.1. and therefore a better measurement is obtained using the MCELS H$\alpha$ image including the continuum.

\begin{figure*}
\includegraphics[width=0.49\linewidth]{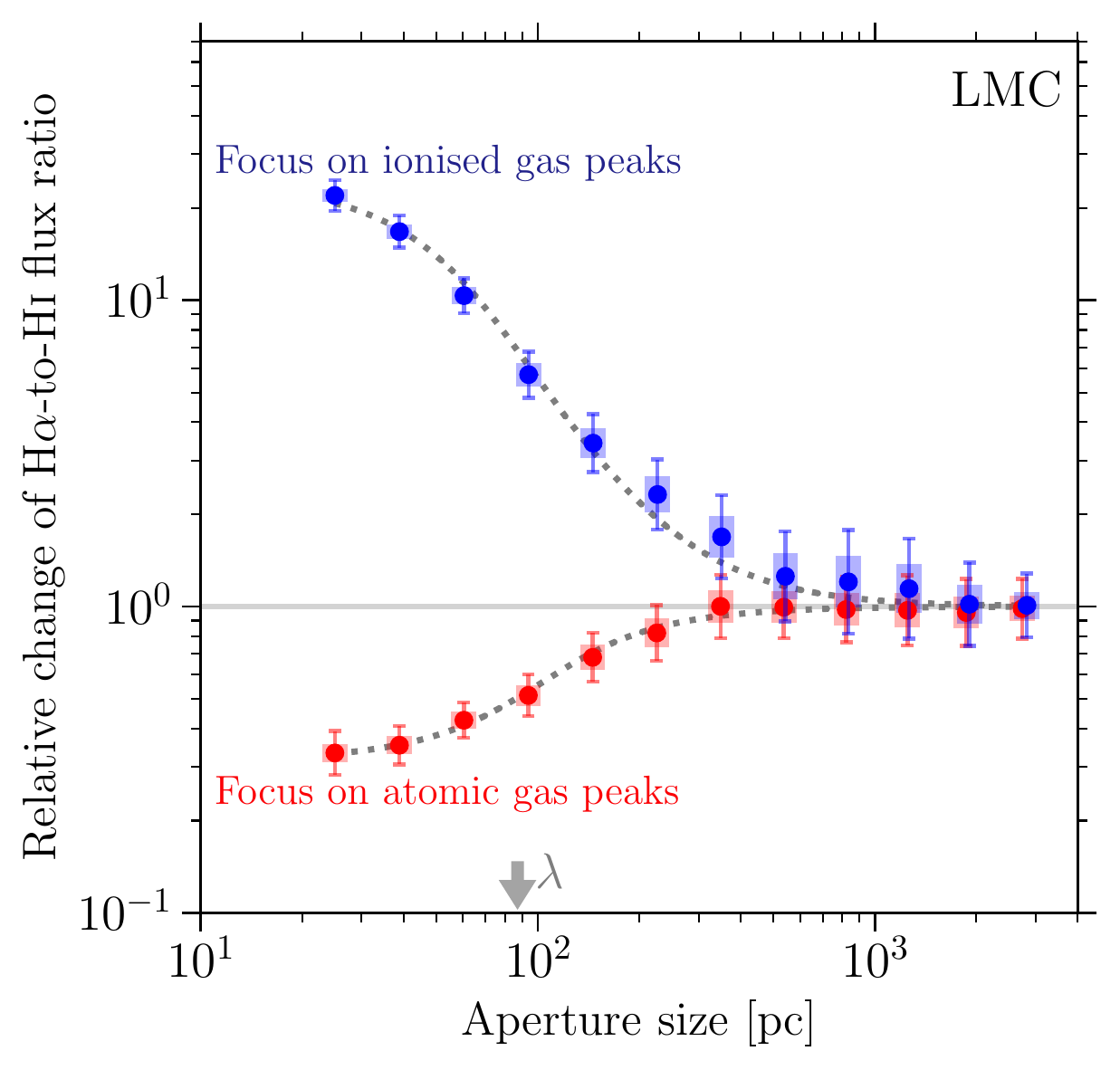}
\includegraphics[width=0.49\linewidth]{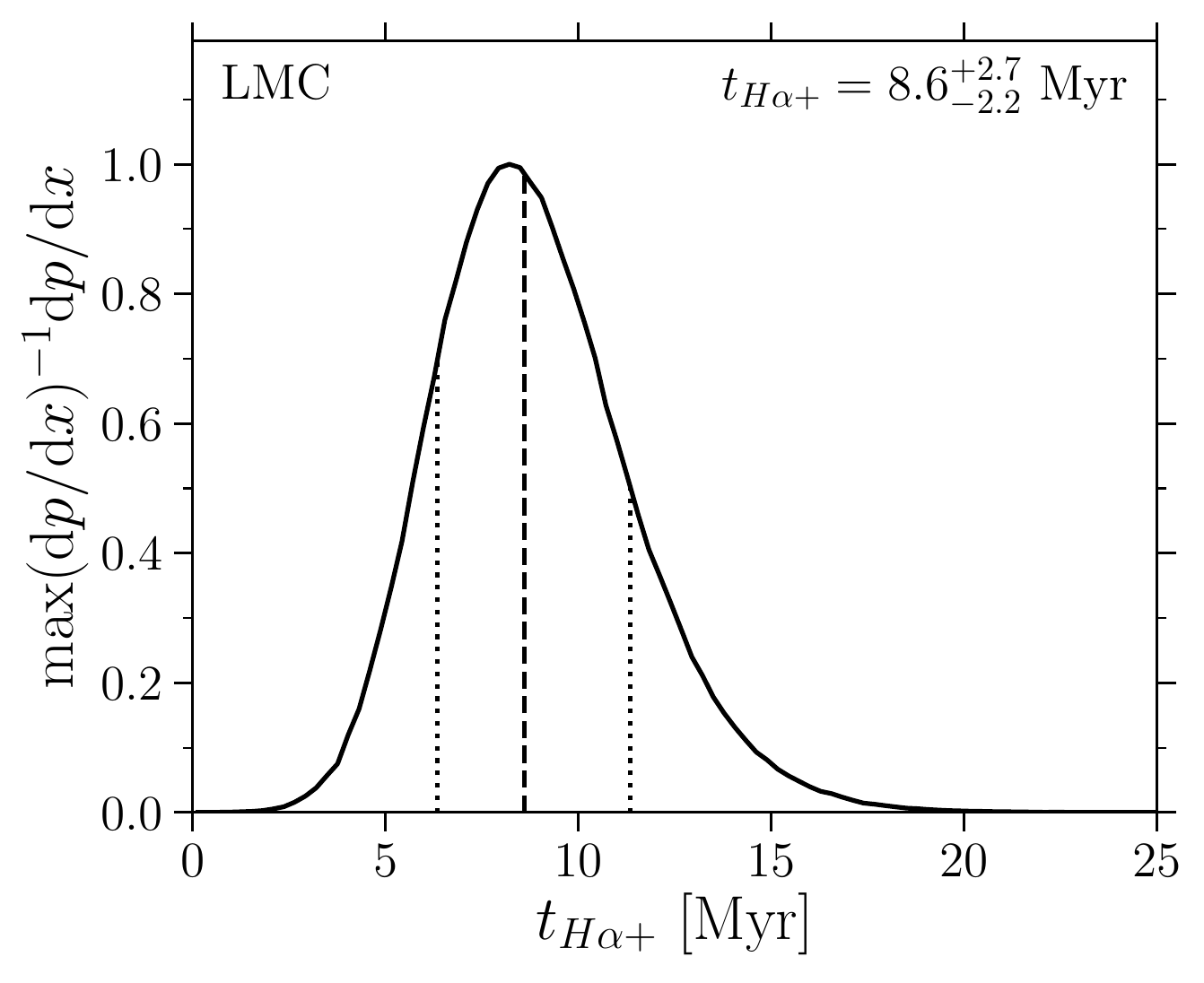}
\caption{\label{test_reverse_fig} The resulting tuning fork and PDF for the derived time-scale for H$\alpha$ plus continuum emission using the previously derived time-scale for H\,{\sc i} as a reference. Note that the derived time-scale is remarkably close to the reference time-scale used in the original H{\sc eisenberg} run.}
\end{figure*}

Finally, we perform a H{\sc eisenberg} analysis using the H\,{\sc i} map as the reference map with an associated time-scale of 47.8\,Myr, to derive the time-scale associated with the MCELS H$\alpha$ emission map.
The resulting H$\alpha$-to-H{\sc i} ratio is shown as a function of aperture size in Fig. \ref{test_reverse_fig}, alongside the PDF for the derived H$\alpha +$ time-scale. The derived time-scale is $8.6\substack{+2.7\\-2.2}$\,Myr, entirely consistent with the calculated reference time-scale of $8.5\substack{+1.0\\-0.8}$\,Myr.

By performing the tests presented in this appendix, we effectively \textquoteleft chain together\textquoteright different pairs of tracers. This application of our methodology demonstrates that the process of applying the uncertainty principle to observational data is both commutative and transitive. It therefore provides an important observational validation of the reference timescales derived by \citet{Haydon2018}.

%%%%%%%%%%%%%%%%%%%%%%%%%%%%%%%%%%%%%%%%%%%%%%%%%%

% Don't change these lines
\bsp	% typesetting comment
\label{lastpage}
\end{document}